\documentclass[letterpaper]{article}

\usepackage[american]{babel}
\usepackage{csquotes}
\usepackage[style=apa]{biblatex}
\usepackage{commath}
\usepackage{amssymb}
\usepackage{amsmath}
\usepackage{bbm}
\usepackage{commath}
\usepackage{graphicx}
\usepackage{xcolor}
\usepackage{import}
\usepackage{subcaption}
\usepackage{placeins}
\usepackage[hidelinks]{hyperref}
\usepackage[noend,noline,ruled]{algorithm2e}
\usepackage{todonotes}
\usepackage{setspace}
\usepackage[misc]{ifsym}
\def\cb{{\mathrm{b}}}
\def\ck{{\mathrm{k}}}
\def\mass{{\mathrm{M}}}

\newcommand\cat[1]{C_{#1}}
\newcommand{\me}{\mathrm{e}}
\def\definition{{\overset{\operatorname{def}}{=}}}
\newcommand{\sv}{\mathbf{x}}

\newcommand{\condensation}{\Pi} 
\newcommand{\cleavage}{X} 
\newcommand{\reduction}{A} 

\def\cc{{\mathrm{c}}}
\def\clcoratio{{\frac{\cc_\cleavage}{\cc_\condensation}}}

\newcommand\reducer{\Rightarrow}
\newcommand\reducesto{\Rightarrow^*}

\newcommand*{\ldblbrace}{\left\{\mskip-5mu\left\{}
\newcommand*{\rdblbrace}{\right\}\mskip-5mu\right\}}

\graphicspath{{figures/}}

\newcommand{\citet}{\textcite}
\renewcommand{\cite}{\textcite}
\newcommand{\citep}{\parencites}

\newcommand{\ifinline}[1]{\ifdefined\journal\else#1\fi}

\newcommand\blfootnote[1]{%
  \begingroup
  \renewcommand\thefootnote{}\footnote{#1}%
  \addtocounter{footnote}{-1}%
  \endgroup
}

\title{Emergence of Self-Reproducing Metabolisms as Recursive Algorithms in an Artificial Chemistry}
\author{Germán Kruszewski$^1$\textsuperscript{\Letter}\blfootnote{\Letter~Corresponding Author}\\{\small \texttt{german.kruszewski@naverlabs.com}} \and Tomas Mikolov$^2$\\{\small \texttt{tmikolov@gmail.com}}}
\date{$^1$ \small Naver Labs Europe, Grenoble, France\\%
    $^2$ CIIRC CTU, Prague, Czech Republic }

\newcommand{\keywords}[1]{\textbf{Keywords:} #1.}

\ifdefined\journal
\fi
\begin{document}
\maketitle
\begin{abstract}
    One of the main goals of Artificial Life is to research the conditions 
    for the emergence of life, not necessarily as it
    is, but as it could be.
    Artificial Chemistries are one of the most important tools for this
    purpose because they provide us with a basic framework to investigate
    under which conditions metabolisms capable of reproducing themselves, and
    ultimately, of evolving, can emerge.
    While there have been successful attempts at producing examples of emergent self-reproducing metabolisms, the set of rules involved remain too complex to shed much light on the underlying principles at work. 
    In this paper, we hypothesize that the key property needed for self-reproducing metabolisms to emerge is the existence of an auto-catalyzed subset of Turing-complete reactions.
    We validate this hypothesis with a minimalistic Artificial Chemistry with conservation laws, which is based on a Turing-complete rewriting system called Combinatory Logic.
    Our experiments show that a single run of this chemistry, starting from a tabula rasa state, discovers – with no external intervention – a wide range of emergent structures including ones that self-reproduce in each cycle.
    All of these structures take the form of recursive algorithms that acquire basic constituents from the environment and decompose them in a process that is remarkably similar to biological metabolisms.
\end{abstract}


\newcommand{\smalltt}[1]{\texttt{\footnotesize #1}}
\newcommand{\algreactor}{
\begin{algorithm}[H]
    \SetInd{0em}{1.1em}
    \DontPrintSemicolon
    \KwIn{Num. of combinators $N_I$, $N_K$, $N_S$; Reaction rates $k_\cleavage$, $k_\condensation$, $k_\reduction$}
    \SetKw{KwTrue}{True}
    \SetFuncSty{smalltt}
    \SetNoFillComment
    \SetKwComment{Comment}{$\triangleright$\ }{}
    Initialize multiset $\mathcal{P} \gets \ldblbrace I: N_I , K: N_K , S: N_S \rdblbrace$\;
    Initialize time $t \gets 0$\;
    \While{\KwTrue}{
        $\sv \gets \mathcal{P}$\;
        Compute $a_\cleavage(\sv), a_\condensation(\sv), a_\reduction(\sv)$, and $a_0(\sv)$  following Equations \ref{eq:propensity_start}-\ref{eq:propensity_end}\;
        Sample reaction type $J \in \{\cleavage, \condensation, \reduction\}$ with prob.  $p(J) = a_J(\sv)/a_0(\sv)$ \;
        \If(\tcp*[h]{cleave}){$J = \cleavage$}{
            Sample expr. $x$ w/prob. $p(x) = \frac{\mathbbm{1}\left[|x|>1\right]k_\cleavage \sv_x }{a_\cleavage(\sv)}$\;
            $j \gets \left[x \rightarrow x_1 + x_2 \right]$ where $x = (x_1 x_2)$\;
        }
        \ElseIf(\tcp*[h]{condense}){$J = \condensation$}{
            Sample expr. $x_1$ w/prob. $p(x) = \sv_{x} / \lVert\sv\rVert_1$\;
            Sample expr. $x_2$ w/prob. $p(x) = \frac{\mathbbm{1}\left[x=x_1\right](\sv_{x}-1) + \mathbbm{1}\left[x\neq x_1\right]\sv_{x}}{ (\lVert\sv\rVert_1 - 1)}$\;
            $j \gets \left[x_1 + x_2 \rightarrow (x_1 x_2) \right]$
        }
        \ElseIf(\tcp*[h]{reduce}){$J = \reduction$}{
            Sample reaction $j$ w/prob. $p(j) = a_j(\sv) / a_\reduction(\sv)$ where
            $a_j(\sv) = k_\reduction \left(\mathbbm{1}\left[j \in A_I\right] \sv_{j_1} + \mathbbm{1}\left[j \in A_K\right] \sv_{j_1} + \mathbbm{1}\left[j \in A_S\right] \sv_{j_1} \sv_{j_2}\right)$ and
            $A_{Z \in \{S, K, I\}}$ are all $Z$-reductions following Rules \ref{eq:red0}-\ref{eq:redn}\;
        }
        Let $x_1,\dots,x_n,y_1,\dots,y_m $ s.t. $j = x_1 + \dots + x_n \rightarrow y_1 + \dots +y_m$\;
        $\mathcal{P} \gets \{\{x: \sv_x - \sum_{i=1}^n \mathbbm{1}[x = x_i] + \sum_{i=1}^m \mathbbm{1}[x = y_i]\}\}$\;
        $t \gets t + a_0^{-1}(\sv)\ln(u^{-1})$ where $u \sim \mathcal{U}(0, 1)$\;
    }
    \caption{Time Evolution Algorithm}\label{al:reactor}
\end{algorithm}
}

\newcommand{\figquine}{
\begin{figure}[htpb]
\centering{
    \resizebox{240pt}{!}{
    \def\svgwidth{350pt}
    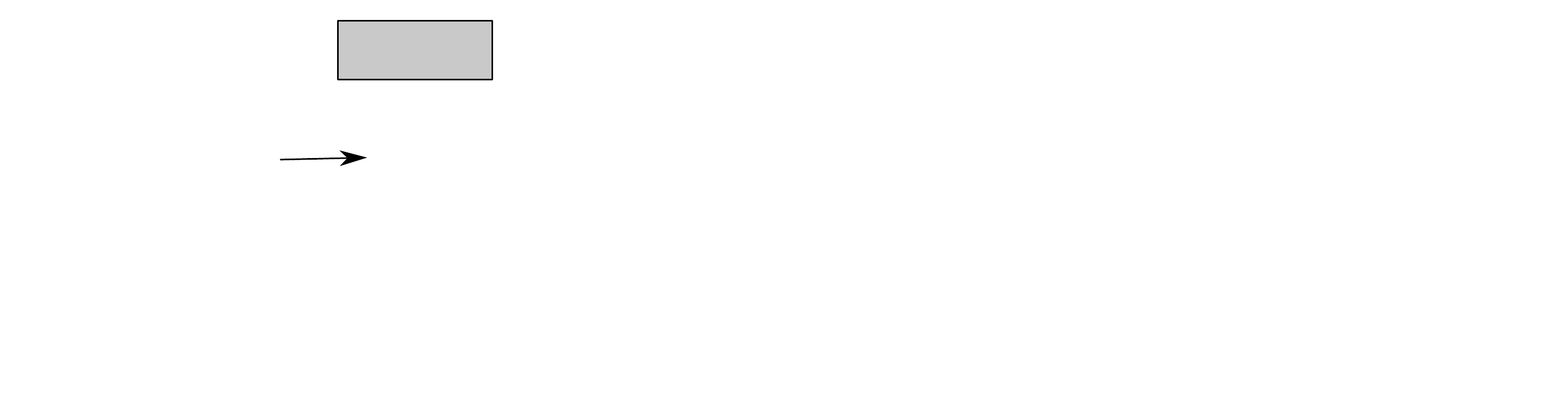
    }
}
\caption{$r_1$--$r_5$ form an autocatalytic
set, granted that $(SII)$ belongs to the food set. $(SII(SII))$'s metabolic cycle
starts with $r_1$ reducing the $S$ combinator, while taking $(SII)$ as
reactant.
Then, the cycle is completed by the reduction of the two identity combinators,
in any of the possible orders.} 
\label{fig:autocatalytic_set}
\end{figure}
}

\newcommand{\figrecursive}{
\begin{figure}[htb]
    \centering
    \resizebox{170pt}{!}{
    \def\svgwidth{240pt}
    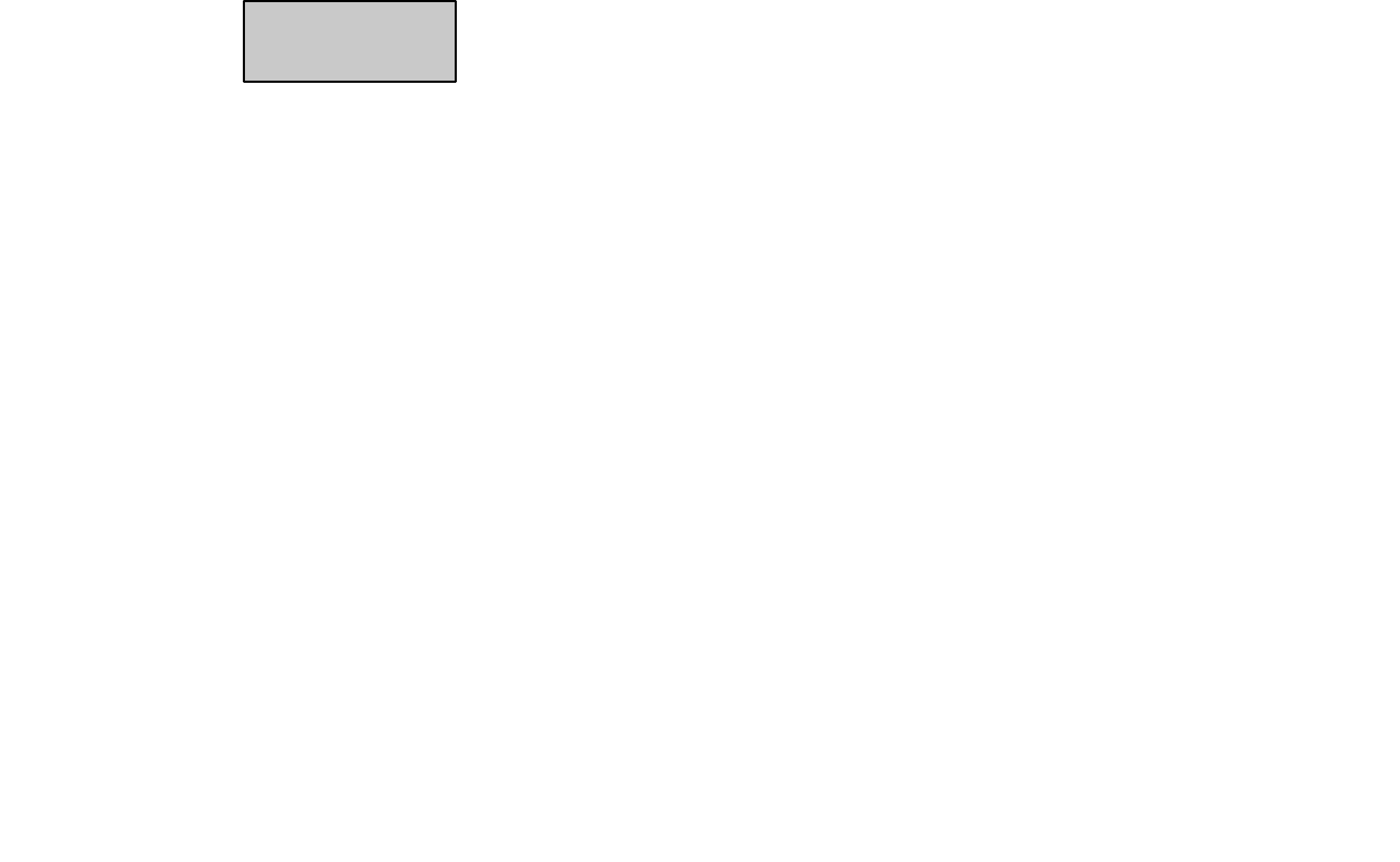
    }
\caption{One of the possible pathways in the reduction of the tail-recursive structure
$(AA)$ with $A=(S(SI)I)$. It appends one $A$ to itself by metabolizing another copy
absorbed from the environment.} 
\label{fig:recursive}
\end{figure}
}

\newcommand{\figselfrepro}{
\begin{figure}[htpb]
\centering{
    \resizebox{200pt}{!}{
    \def\svgwidth{250pt}
    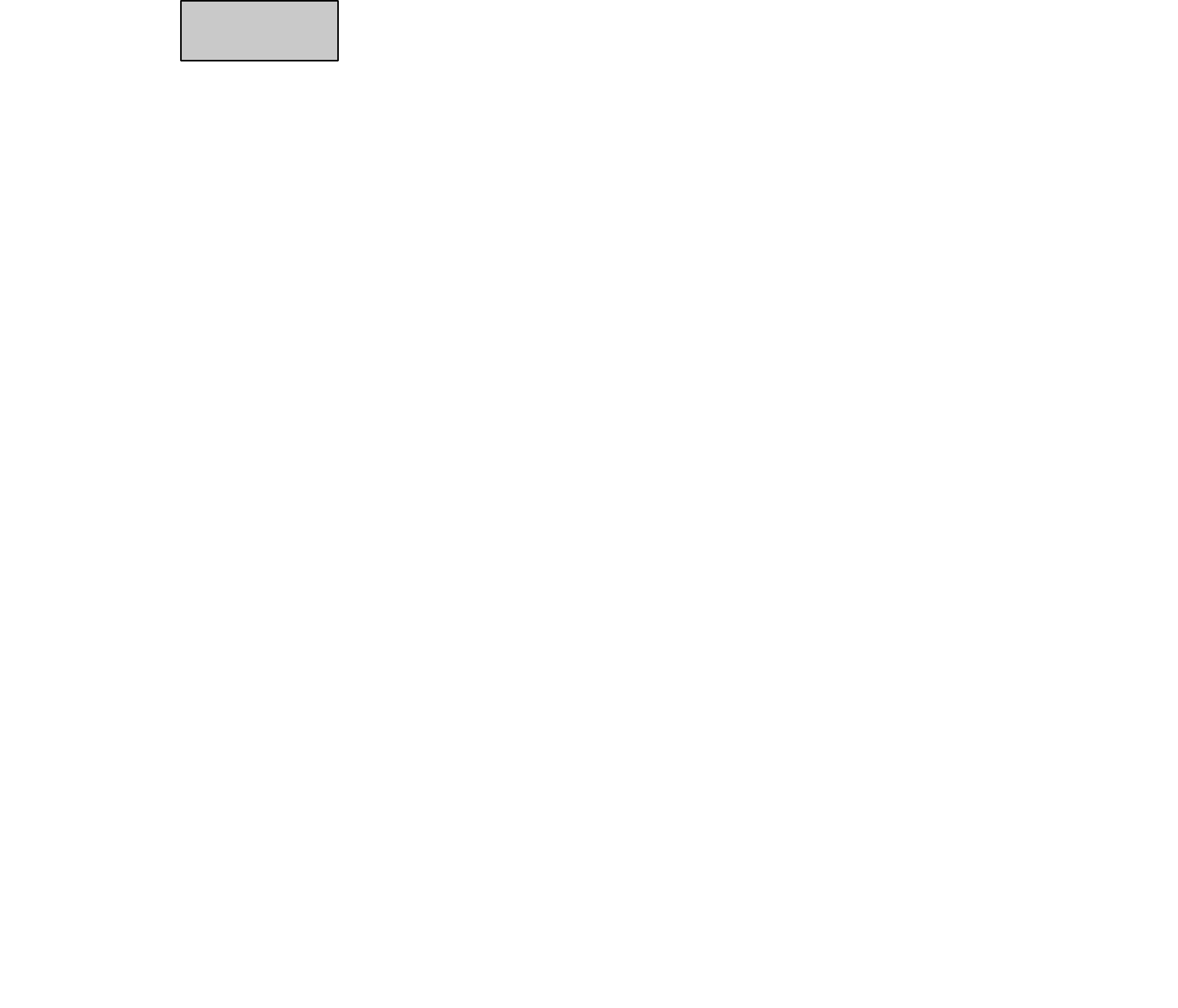
    }
}
\caption{Metabolic cycle (showing one of the possible pathways) of a self-reproducing structure that emerges from the dynamics of Combinatory Chemistry.
    Starting from $(AA)$, where $A=(SI(S(SK)I))$, it acquires three copies of
    $A$ from its environment and uses two to create a copy of itself,
    metabolizing the third one to carry out the process.
}
\label{fig:self-reproducing}
\end{figure}
}

\newcommand{\figweighting}{
\begin{figure}[htpb]
    \begin{subfigure}[t]{0.48\textwidth}
        \centering
        \includegraphics[width=\linewidth]{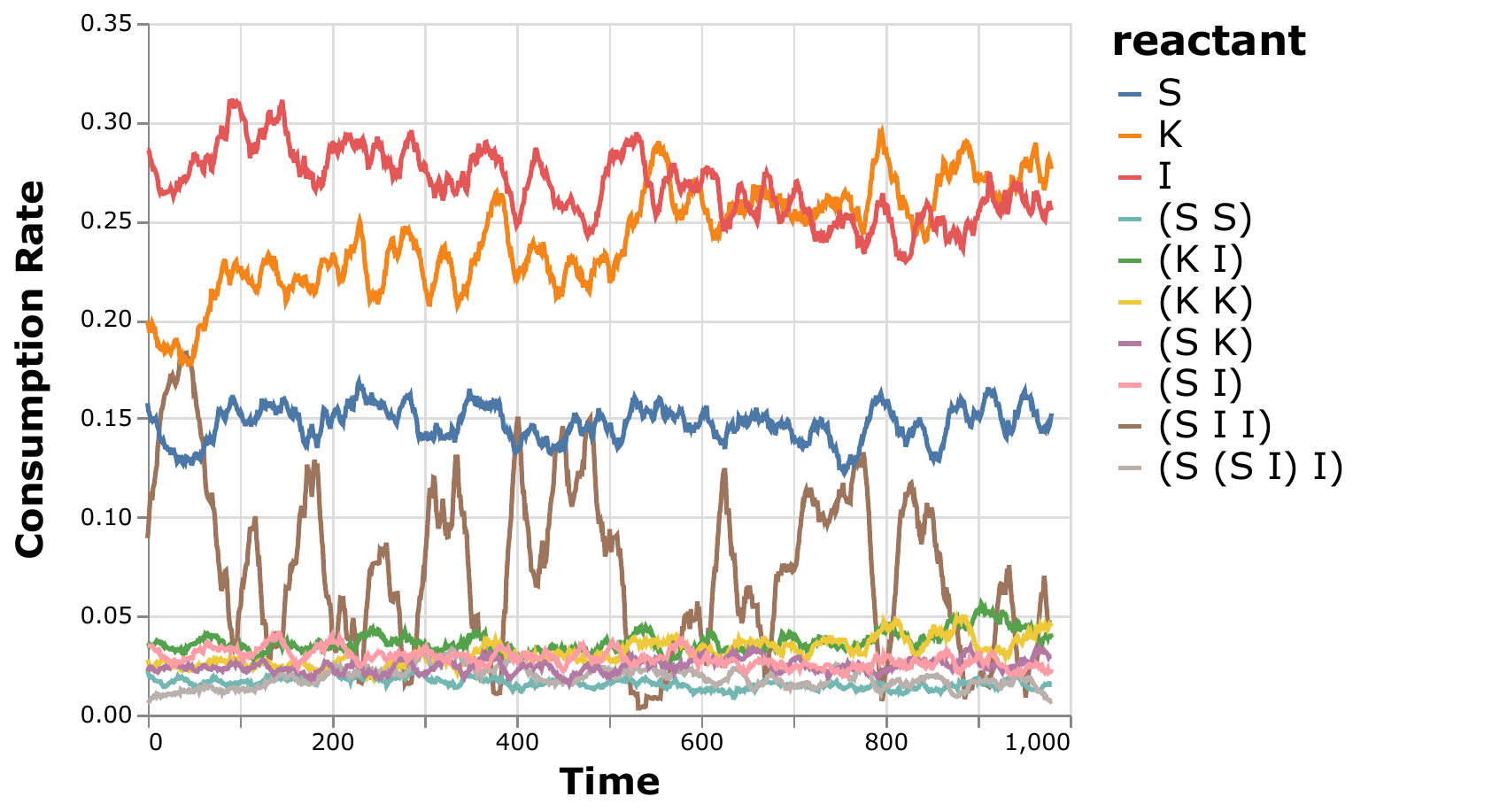}
        \caption{Reactant consumption $C(x)$}
        \label{fig:consumption-rates-base}
    \end{subfigure}\hfill
    \begin{subfigure}[t]{0.48\textwidth}
        \centering
        \includegraphics[width=\linewidth]{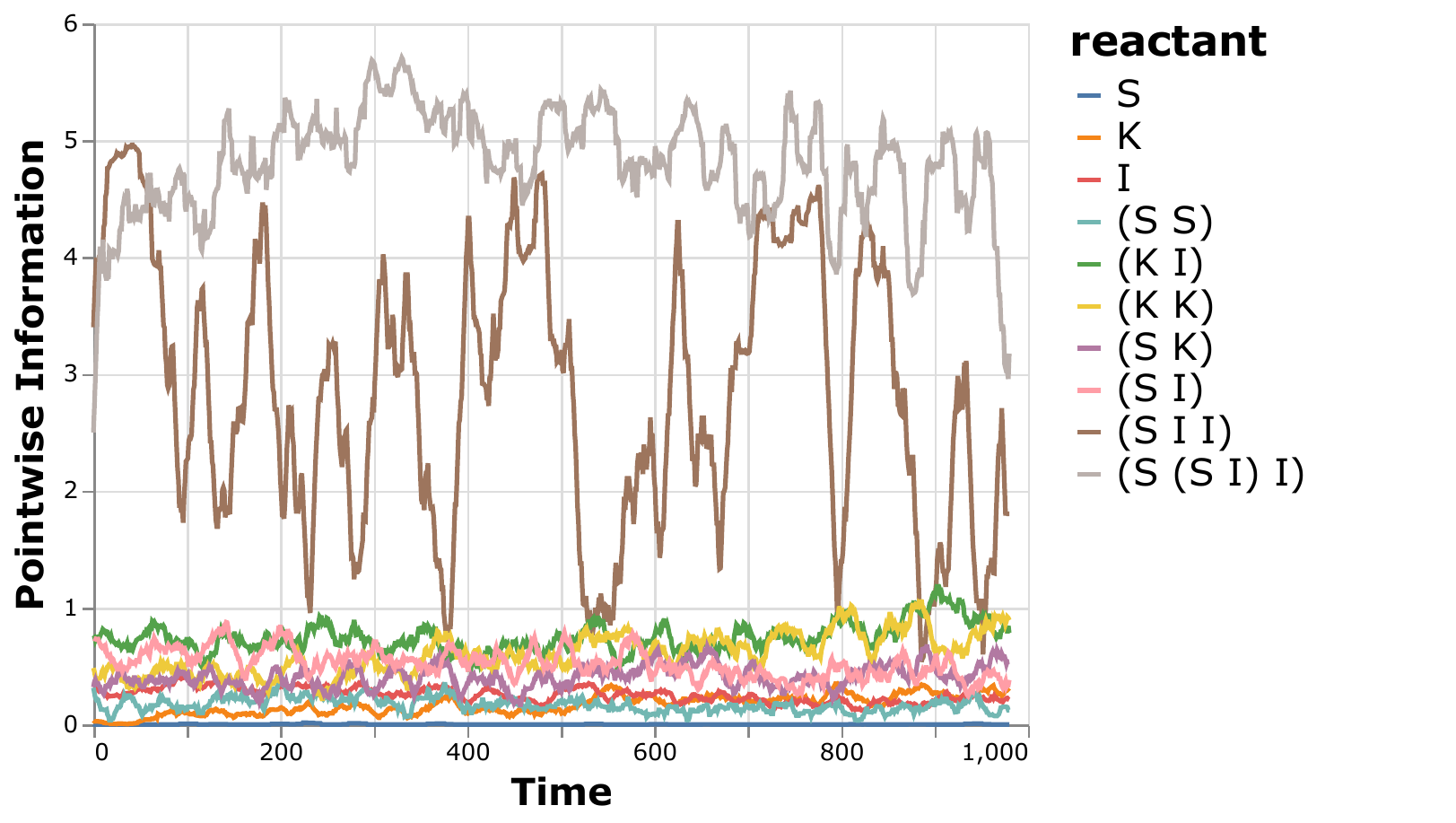}
        \caption{Pointwise Information $I(x)$}
        \label{fig:consumption-rates-info}
    \end{subfigure}
    \caption{Reactant consumption metrics for the most consumed reactants $x$
        in a simulation with $10$k combinators.}
\end{figure}
}

\newcommand{\figglobalmetrics}{
\begin{figure*}[hbt]
    \centering
    \begin{subfigure}[t]{0.3\textwidth}
        \includegraphics[width=\linewidth]{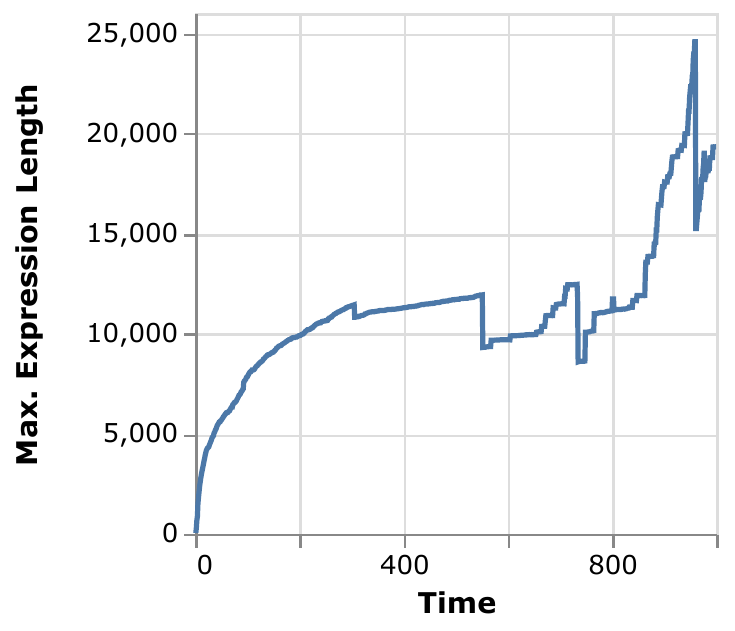}
        \caption{Max. expression length}
        \label{fig:max-length}
    \end{subfigure}
    \begin{subfigure}[t]{0.3\textwidth}
        \includegraphics[width=\linewidth]{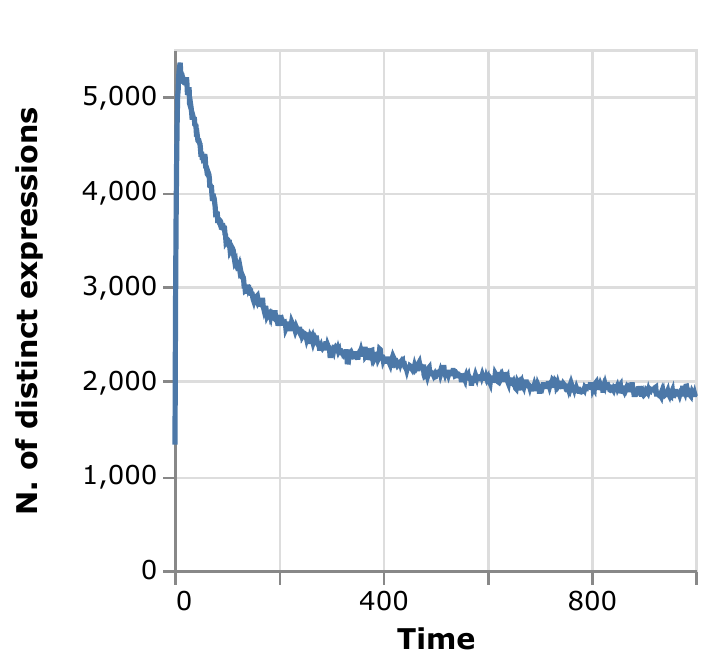}
        \caption{Diversity}
        \label{fig:diversity}
    \end{subfigure}
    \begin{subfigure}[t]{0.3\textwidth}
        \includegraphics[width=\linewidth]{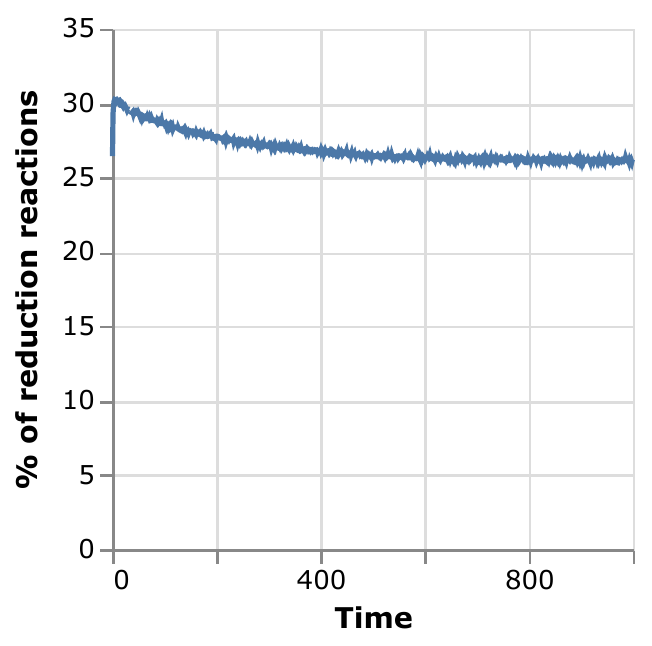}
        \caption{Reduction Probability}
        \label{fig:p-reduce}
    \end{subfigure}
    \caption{Global metrics for a system initialized with $1$M uniformly distributed $S$, $K$, and $I$ combinators.}
\end{figure*}
}

\newcommand{\figemergence}{
\begin{figure*}[hbt]
    \centering
    \begin{subfigure}[t]{0.48\textwidth}
        \includegraphics[width=\linewidth]{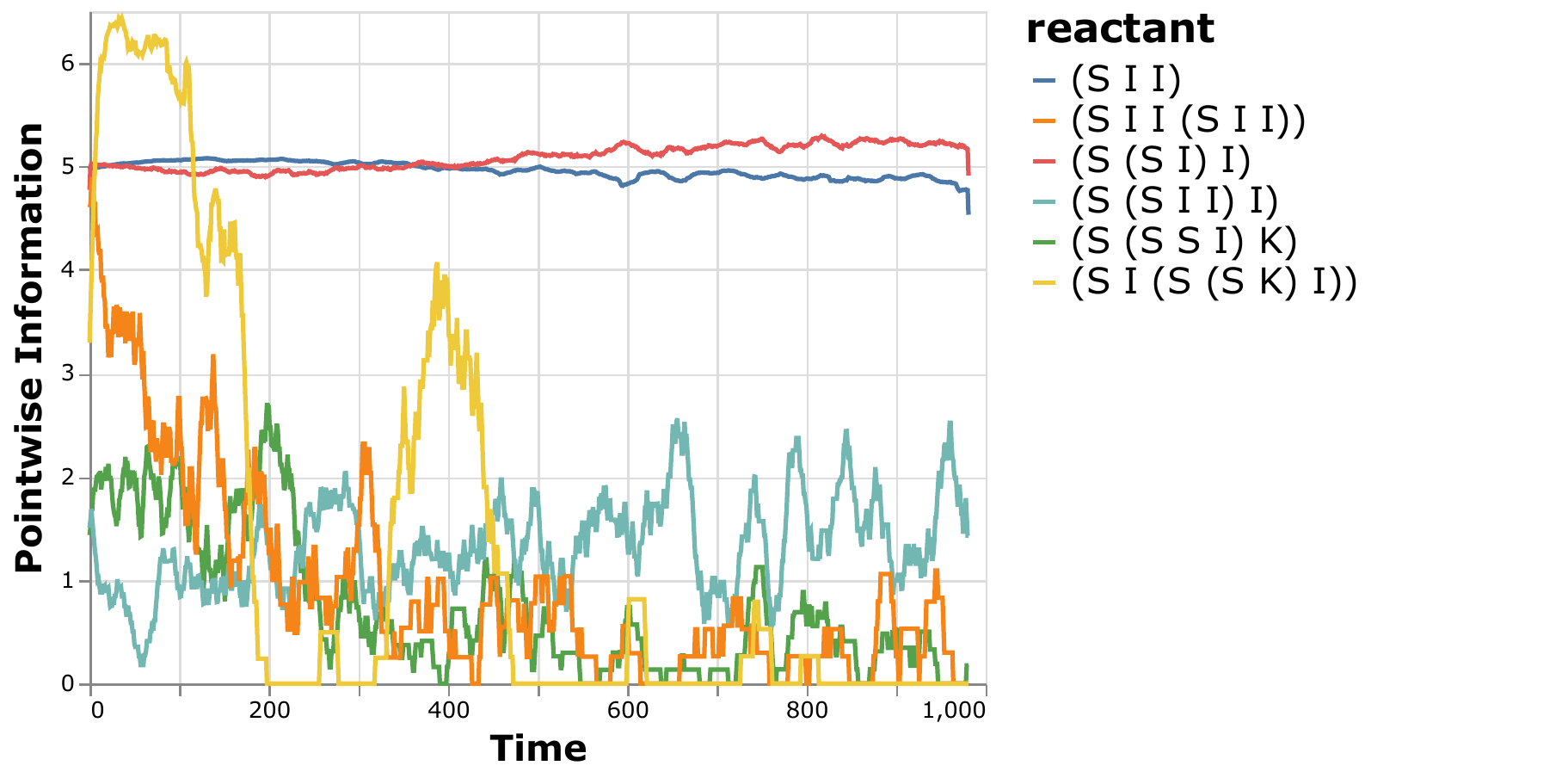}
        \caption{Pointwise information.}
        \label{fig:consumption}
    \end{subfigure}\hfill
    \begin{subfigure}[t]{0.48\textwidth}
        \includegraphics[width=\linewidth]{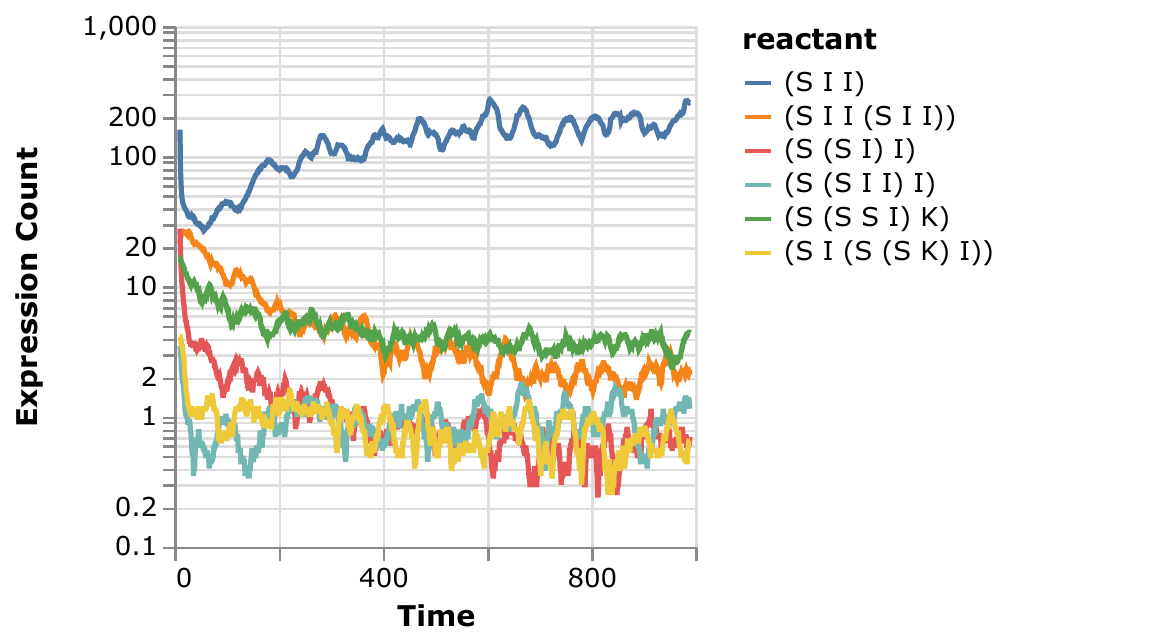}
        \caption{Multiplicity in $\mathcal{P}$}
        \label{fig:reactants-count}
    \end{subfigure}
    \caption{Simulation results for $1$M uniformly selected $S$, $K$ and $I$ combinators
    for $1000$ generations, on a set of manually chosen reactants.}
\end{figure*}
}

\newcommand{\figotherbases}{
\begin{figure}[htpb]
    \centering
    \begin{subfigure}[t]{0.48\textwidth}
    \centering
    \includegraphics[width=0.8\linewidth]{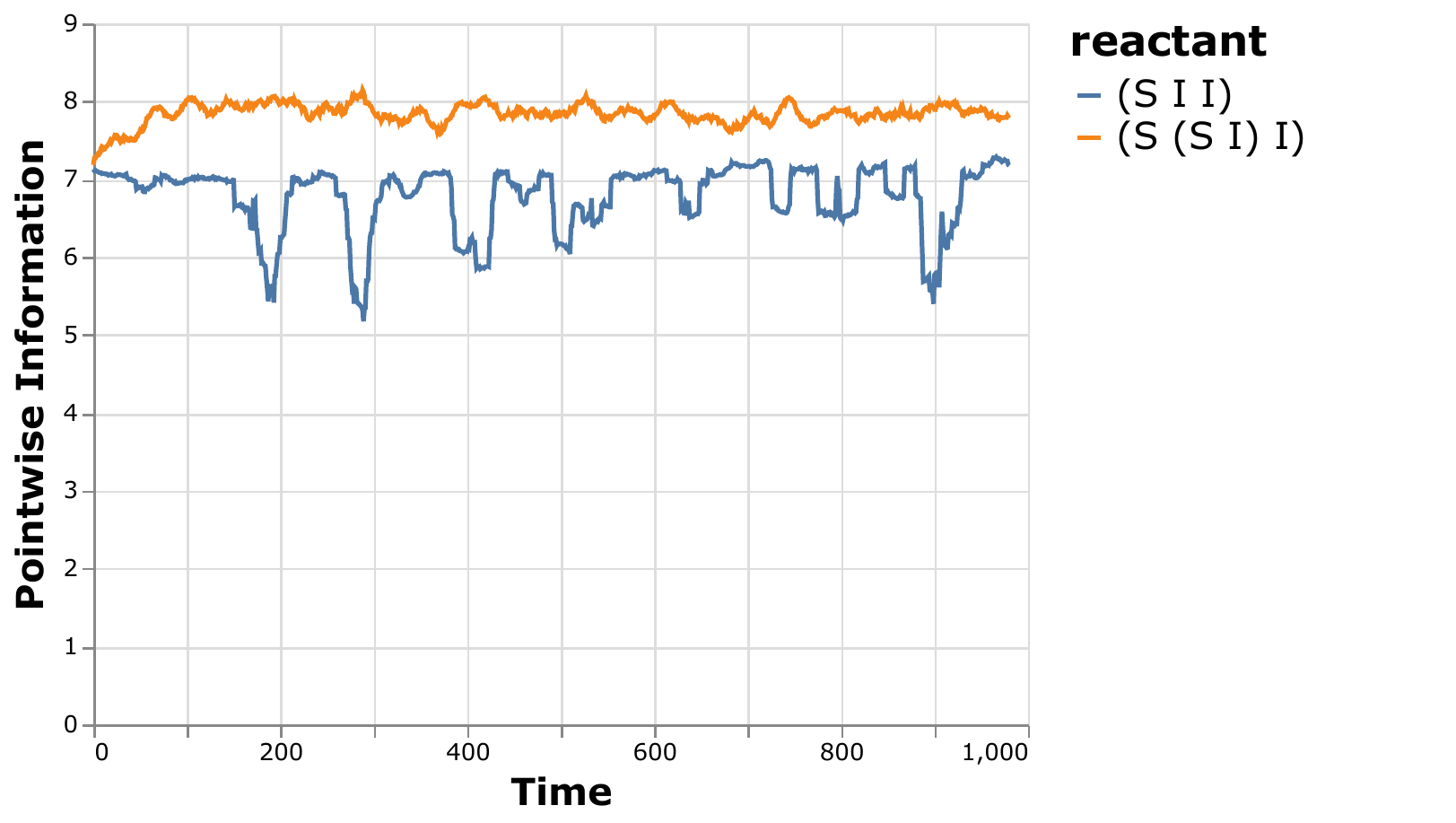}
    \caption{$10$k uniformly distributed $S$ and $I$ combinators.}
    \label{fig:base-SI}
    \end{subfigure}\hfill
    \begin{subfigure}[t]{0.48\textwidth}
    \centering
    \includegraphics[width=0.8\linewidth]{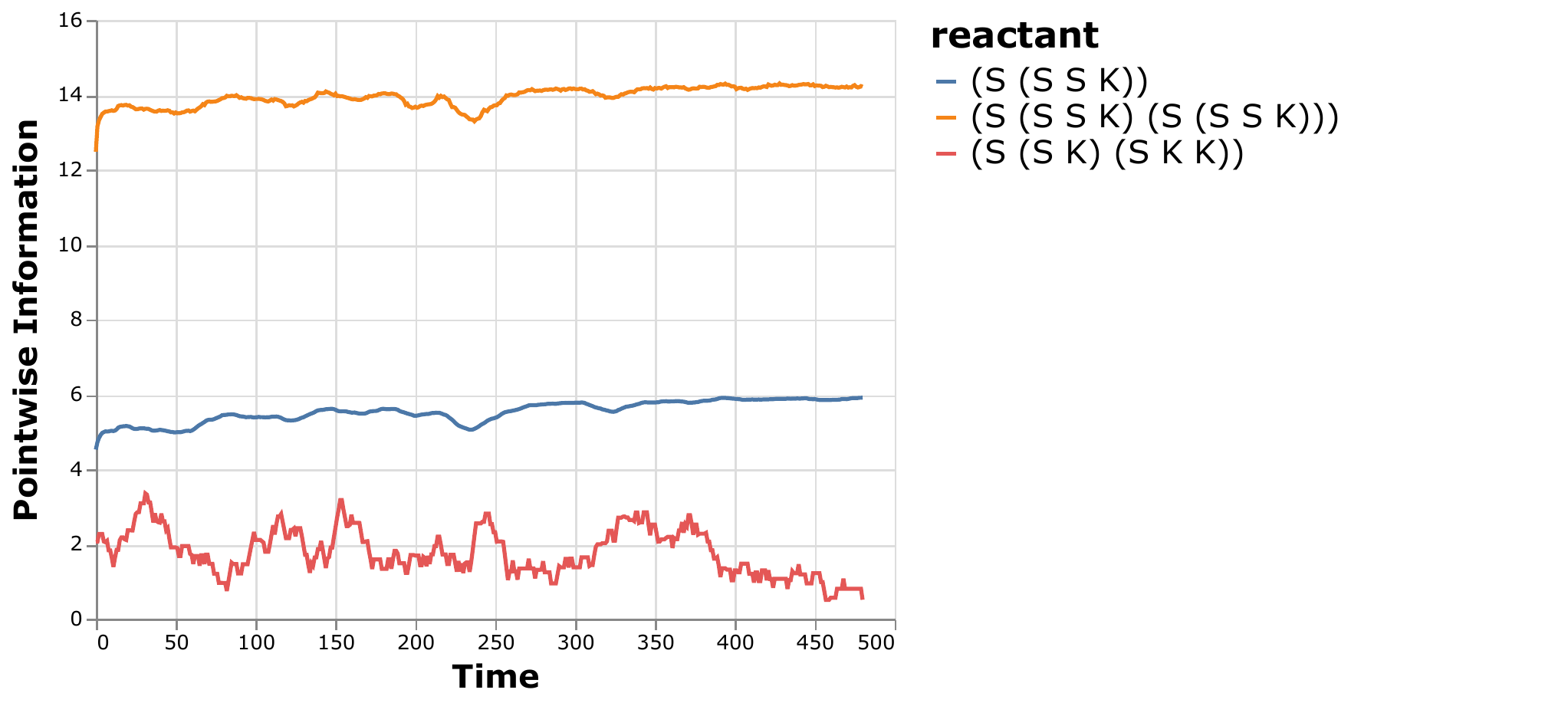}
    \caption{$1$M uniformly distributed $S$ and $K$ combinators.}
    \label{fig:base-SK}
    \end{subfigure}
    \caption{Pointwise Information for a selected set of reactants on different combinator bases.}
\end{figure}}

\section{Introduction}

One central area of focus for Artificial Life is to characterize the conditions that lead to the emergence of living systems. 
More precisely, the goal is to understand under which conditions metabolisms capable of sustaining themselves in time, reproducing, and, ultimately, of evolving, can emerge.
Artificial Chemistries can be used to reveal this process by simulating the properties of natural chemical systems at different levels of abstraction \parencites[see][for a thorough review]{Dittrich:2001}. The driving hypothesis is that complex organizations emerge from the interactions of simpler components thanks to self-organizing attractors in chemical networks \citep{Walker:1966, Wuensche:etal:1992, Kauffman:1993}.
While some Artificial Chemistries model as closely as possible the properties of the chemistry that gave rise to life on Earth \citep{Flamm:etal:2010, Hogerl:2010, Young:Neshatian:2013}, others leave out the particularities of natural chemistry to focus only on their hypothesized core computational properties \citep{Fontana:1994, Speroni:2000,
Hutton:2002,Tominaga:etal:2007,Buliga:Kauffman:2014,Sayama:2018}. Interestingly, some of these studies have described systems that produce emergent metabolic structures \citep{Bagley:Farmer:1992}, and some others feature self-reproducing structures as well \citep{Hutton:2002,Young:Neshatian:2013}, at times also capable of undergoing evolution~\citep{Hutton:2007}. However, it is still not clear which of the properties in these chemical systems are central to the emergence of these structures. Yet, gaining such insights is a crucial step for deriving more general biological theories grounded both in life-as-it-is and life-as-it-might-be ~\citep{Langton:1994}. 

Here, we hypothesize that self-reproducing metabolisms emerge as a subset of auto-catalyzed reactions within a Turing-complete set. To validate this idea, we introduce Combinatory Chemistry, an Artificial Chemistry designed to capture the core computational properties of living systems. It consists in a set of auto-catalyzed reactions, which are based on the rewriting rules of Combinatory Logic~\citep{Schonfinkel:1924, Curry:1958}. These reactions inherit from the original rewiring rules the capacity to perform universal computation. Furthermore, we adapted the rules so that the ensuing reactions would have conservative dynamics. Completing the set of possible reactions, there are random mixing rules that act at a far lower rate. 

The resulting chemical system is strongly constructive~\citep{Fontana:Buss:1993}, which means that as the system evolves in time it can create --by chance or through the action of its own components-- new components that can in turn modify its global dynamics. Furthermore, thanks to its universal computation capabilities, there is no theoretical limit to the complexity that emergent forms can have. On the other hand, because of the conservation dynamics, the memory cost of the system remains constant without needing to introduce external perturbations (such as randomly removing elements from the system), while possibly also providing a natural source of selective pressure.

In contrast to previous work that explicitly banned expressions which would not reduce to a normal form~\citep{Fontana:1994,Speroni:2000}, Combinatory Chemistry can handle them adequately by distributing their (potentially infinite) computation steps over time as individual reactions. Also, with respect to an earlier version of this work \citep{Kruszewski:Mikolov:2020}, here we have further simplified the system, dropping the “reactant assemblage” mechanism through which we used to feed emergent structures with their required resources. This mechanism enabled them to grow their activity and thus be more easily spotted, but it also biased the evolution of the system. Instead, here we have devised a new metric to identify these structures in naturally occurring resource conditions. Moreover, we use Gillespie's algorithm~\citep{Gillespie:1977} to simulate the time evolution of the system to obtain unbiased samples of the proposed process, instead of a more simplistic algorithm that was used before.

Starting from a tabula rasa state consisting of only elementary components, Combinatory Chemistry produces a diversity explosion, which then develops into a state dominated by self-organized emergent autopoietic structures~\citep{Varela:Maturana:1973}, including recursively growing and self-replicating ones. Notably, all these types of structures emerge during a single run of the system without requiring any external intervention. Furthermore, they preserve themselves in time by absorbing compounds from their environment and decomposing them step by step, in a process that has a striking resemblance with the metabolism of biological organisms. These structures take the form of recursive algorithms that never reach a normal form (i.e., halting point) as long as sufficient resources are present in their environment. Notably, we found that Turing-completeness was required in Combinatory Chemistry for representing self-replicating structures, but not for simple autopoietic or recursively growing ones.

This paper is organized as follows. First, we describe earlier work in Artificial Chemistry that is most related to the approach introduced here. Then, we explain the basic workings of Combinatory Logic and how we adapted it into an Artificial Chemistry. Third, following earlier work, we discuss how autocatalytic sets can be used to detect emerging phenomena in this system and propose a novel measure of emergent complexity, which is well adapted to the introduced system. Finally, we describe our experiments showcasing the emergence of complex structures in Combinatory Chemistry and discuss their implications.

\section{Artificial Chemistries}
Artificial Chemistries (AC) are models inspired in natural chemical systems
that are usually defined by three different components: a set of possible
molecules, a set of reactions, and a reactor algorithm describing the reaction
vessel and how molecules interact with each other~\citep{Dittrich:2001}.
In the following discussion, we will focus on the algorithmic chemistries that are
the closest to the present work.

AlChemy~\citep{Fontana:1994} is an AC where molecules are given by
$\lambda$-calculus expressions. 
$\lambda$-calculus is a mathematical formalism that, like Turing machines, can
describe any computable function. 
In AlChemy, pairs of randomly sampled expressions are joined through function
application, evaluated, and the corresponding result is added back to the population.
To keep the population size bounded, expressions are randomly discarded.
Fontana and Buss showed that expressions that computed themselves quickly
emerged in this system, which they called level 0 organizations.
Furthermore, when these expressions were explicitly prohibited, a more complex
organization emerged where every expression in a set was computed by
other expressions within the same set (level 1 organizations).
Finally, mixing level 1 organizations could lead to higher order interactions
between them (level 2 organizations).
Yet, this system had some limitations.
First, each level of organization was only reached after external
interventions.
In addition, programs were evaluated using $\beta$-reductions, which require that they
reach a normal form, namely, that there are no more $\lambda$-calculus rules
than can be applied.
Thus, recursive programs, which never reach a normal form, are banned from the
system.
Here, we use weak reductions instead, allowing the system to 
compute the time evolution of programs that never reach a formal form. 
Interestingly, it is exactly in this way that emergent metabolisms are represented.
Furthermore, in AlChemy, two processes were introduced as analogues of food and waste, 
respectively.
First, when expressions are combined, they are not removed from the
system, allowing the system to temporarily grow in size.
Second, expressions which after being combined with existing expressions
do not match any $\lambda$-calculus reduction rules are removed.
Without these processes, complex organizations fail to emerge.
Yet, it is not clear under which circumstances these external interventions
would not be needed anymore in order for the system to evolve autonomously.
Finally, bounding the total number of expressions by randomly removing excess
ones creates perturbations to the system that can arbitrarily affect the
dynamics.
\cite{Fontana:Buss:1996} later proposed MC2, a chemistry based on Linear Logic
that addressed some of these limitations (notably, conservation of mass),
although we are not aware of empirical work on it.

Here, we propose an AC based on Combinatory Logic.
This formalism has been explored before in the context of AC by
\cite{Speroni:2000}.
This work also introduces conservation laws, even though they rely on decomposing
expressions to their individual components, introducing some randomness
into the dynamics that here we have avoided.
Furthermore, like AlChemy, it reduces expressions until they reach their normal
forms, explicitly forbidding recursive and other types of expressions that do
not converge.

Other very related Artificial Chemistries are based on graph rewriting systems.
Squirm3~\citep{Hutton:2002} is a chemistry in which atoms are placed in a 2D
grid where they can react with each other, creating or breaking bonds. 
Interestingly, \citet{Hutton:2002} shows that self-reproducing evolvable chains
can emerge in this environment when using the right set of reactions, which
like in the AC here introduced, they have intrinsic conservation laws.
Yet, it remains unclear what characteristics of those reactions make it
possible for this emergence to occur.
In this work, we study the hypothesis that self-reproducing metabolisms are
linked to recursive programs expressed through a network of auto-catalysed
reactions endowed with universal computation capabilities.
In a different vein, Chemlambda \citep{Buliga:Kauffman:2014,Buliga:2020} is a
Turing-complete graph rewriting AC that allows the encoding of
$\lambda$-calculus and combinatory logic operators. 
As such, it is complementary in many ways to the system proposed here.
While the original Chemlambda did not consider conservation laws, an extension
called Hapax is currently exploring them.
Yet, emergent phenomena have not yet been explored under this formalism.

\section{Combinatory Logic} 
Combinatory Logic (CL) is a minimalistic computational system that was
independently invented by Moses Sch\"onfinkel, John Von Neumann and Haskell
Curry \citep{Cardone:2006}. 
Other than its relevance to Computability Theory, it has also been applied 
in Cognitive Science as a model for a Language of Thought \citep{Piantadosi:2016}.
One of the main advantages of CL is its formal simplicity while capturing
Turing-complete expressiveness.
In contrast to other mathematical formalisms, such as $\lambda$-calculus, it
dispenses with the notion of variables and all the necessary bookkeeping that 
comes with it.
For instance, a function $f(x) = 1 + x + y$ would be nonsensical, and a
function-generating system based on $\lambda$-calculus would need to have
explicit rules to avoid the formation of such expressions.
Instead, CL expressions are built by composing elementary operators called
combinators. 
Here, we restrict to the $S$, $K$ and $I$ combinators, which form a
Turing-complete basis\footnote{As a matter of fact, $S$ and $K$ suffice because
    $I$ can be written as $(SKK)$. The inclusion of $I$ simply allows to express
    more complex programs 
with shorter expressions.}. 
A CL expression is defined either to be a singleton combinator or, recursively,
given two expressions $x$ and $y$ by the application operation $(x y)$.
It is important to note that, by convention, application is left-associative
and thus, $(xyz)$ and $((xy)z)$ are equivalent.
Given an expression $e$ of the form $e=(\alpha X \beta$) where $X$ is a well-formed
sub-expression and $\alpha$ and $\beta$ are some arbitrary left and right context, 
it can be rewritten in CL, as follows:
\begin{eqnarray*}
    \alpha (I f) \beta &\rhd& \alpha f \beta\\
    \alpha (K f g) \beta &\rhd& \alpha f \beta \\
    \alpha (S f g x) \beta &\rhd& \alpha (f x ( g x )) \beta
\end{eqnarray*}
When ($\alpha X \beta$) matches the left hand side of
any of the rules above, the term $X$ is called a ``reducible expression'' or \emph{redex}.
A single expression can contain multiple redexes.
If no rule is matched, the expression is said to be in normal form.
The application of these rules to rewrite any redex is called a (weak)
reduction.
For example, the expression $(SII(SII))$ could be reduced as follows (underlining
the corresponding redexes being rewritten):
$(\underline{SII(SII)}) \rhd (\underline{I(SII)}(I(SII))) \rhd
(SII(\underline{I(SII)})) \rhd (SII(SII))$. 
Thus, this expression, also known as the omega combinator, reduces
to itself.
We will later see that expressions such as this one will be important for the
self-organizing behavior of the system introduced here.
In contrast, $(SII)$ is not reducible because $S$ requires three arguments.\footnote{Also, $I$  cannot be reduced with $I$ as an argument because $(SII) = ((SI)I)$ and thus, the second $I$ is not an argument for the first $I$ but to $(SI)$.}
Additionally, note that $(I(SII)(I(SII)))$ has two redexes that can be rewritten, namely,
the outermost or the innermost $I$ combinators.
Even though many different evaluation order strategies have been defined
\citep{Pierce:2002}, here we opt for picking a redex at random, both because
this is more natural for a chemical system and to avoid limitations that would
come from following a fixed deterministic evaluation order.

\section{Combinatory Chemistry}

One of our main contributions deals with reformulating these reduction rules as 
reactions in a chemical system with conservation laws. 
For this, we postulate the existence of a multiset of CL expressions $\mathcal{P}$ that
react following reduction rules, plus random condensation and cleavages.
In principle, the application of a reduction rule to any expression involves
removing one copy of the expression from the multiset and adding back the
resulting product of the rule.
Note that if we were to apply plain CL rules to reduce these expressions, the
total number of combinators in the system would not be
preserved~\citep{Lafont:1997}.
First, because the application of a reduction rule always removes the first
combinator in the redex from the resulting expression. 
Second, because while the $K$ combinator discards a part of the expression (the
argument $g$), $S$ duplicates its third argument $x$.
Thus, to make a chemical system with conservation laws, we posit that, on
one hand, reduction
operations can generate one or more by-product expressions.
On the other hand, the reduction rules can be applied to more than one
expression simultaneously.
Therefore, using the $+$ symbol to indicate that multiple expressions are being
rewritten when it appears on the left hand side, or more than one expression is being added
back to the multiset when it is on the right hand side, we define reduce reactions for 
an expression or \emph{substrate} $(\alpha X \beta)$, as follows:
\begin{eqnarray}
    \alpha (I f) \beta &\reducer& \alpha f \beta + I \label{eq:red0}%
    \\
    \alpha (K f g) \beta &\reducer& \alpha f  \beta + g + K \label{eq:redK} %
    \\
    \underbrace{\alpha (S f g x) \beta}_\text{substrate} \underbrace{+ x}_{\text{reactant}} &\reducer& \underbrace{\alpha (f x (g x)) \beta}_\text{product} \underbrace{+ S}_{\text{by-product}}\label{eq:redS}
    \label{eq:redn}
\end{eqnarray}
An expression in Combinatory Chemistry is said to be \emph{reducible} if it
contains a Combinatory Chemistry redex (CC-redex).
A CC-redex is a plain CL redex, except when it involves the reduction of an $S$
combinator, in which case a copy of its third argument $x$ (the
\emph{reactant}) must also be present in the multiset $\mathcal{P}$ for it to
be a redex in Combinatory Chemistry.
For example, the expression $SII(SII)$ is reducible if and only if the third
argument of the combinator $S$, namely $(SII)$, is also present in the set.
When a reduction operation is applied, the redex is rewritten following the
rules of combinatory logic, removing any reactant from $\mathcal{P}$ and adding
back to it the \emph{product} and all \emph{by-products}, as specified 
on the right hand side of the reaction.
The type of combinator being reduced gives name to the reaction.
For instance, the $S$-reaction operating on $SII(SII) + (SII)$ removes these
two elements from $\mathcal{P}$, adding back $I(SII)(I(SII))$ and $S$ to it.
Notably, each of these reduction rules preserves the total number of
combinators in the multiset, intrinsically enforcing conservation laws in this
chemistry.
It is also worth noting that each of these combinator plays different roles
in the creation of novel compounds. 
While $K$-reactions split the expression, decreasing its total size and
complexity, $S$-reactions create larger and possibly more complex expressions
from smaller parts.

In contrast to previous attempts in which expressions were combined and then
reduced to normal form~\citep{Fontana:1994, Speroni:2000}, and thus being forced to
exclude expressions that did not reach a normal form, here each reduce reaction
corresponds to a \emph{single} reduction step that can always be computed.
For this reason, we do not need to take any precautions to avoid recursive
expressions, allowing these interesting programs to form part of our system's
dynamics.

Completing the set of possible reactions in this chemistry, condensations
and cleavages can generate novel expressions through random recombination:
\begin{eqnarray}
    x + y \longleftrightarrow (xy) \label{eq:random}
\end{eqnarray}
Condensations correspond to the application operation between $x$ and $y$, 
whereas cleavages are the inverse. Note that cleaving $(xyz)$ can only result
in $(xy) + z$ because, otherwise, the tree structure would not be preserved.

In Combinatory Chemistry, computation takes precedence. 
This means that reduction reactions must happen at much higher rates than
those of random recombination. 
Over the following section we detail how this happens.

\subsection{Temporal Evolution}

The system is initialized with a \emph{tabula-rasa} state containing only
expressions with a single combinator $S$, $K$ or $I$.
In this way, we can be sure that any emergent diversity is the consequence of
the system's dynamics rather than the outcome of an external intervention.
Then, it evolves by sampling reactions following Gillespie's
algorithm~\citep{Gillespie:1977, Gillespie:2007}.
We note that the time evolution algorithm has changed from an earlier version
of this work~\citep{Kruszewski:Mikolov:2020} in favor of this more principled
approach.

More precisely, we define a propensity function $a_j(\sv)$ for each reaction
$j$, which computes the unnormalized probability that the reaction $j$ will
occur within an infinitesimal time interval given the system's state vector $\sv$.
The component $\sv_x$  indicates the number of instances of an
expression $x$ that are present in $\mathcal{P}$.
To define the propensity functions, we make use of reaction rate constants
$k_\cleavage$, $k_\condensation$, $k_{\reduction_S}$, $k_{\reduction_K}$,
$k_{\reduction_I}$, for the cleavage, condensation, and $S$, $K$, and
$I$-reduction reactions, respectively.
Importantly, because in Combinatory Chemistry computation takes precedence,
reaction rates $k_j$ of reduction reactions must be significantly larger than
those of random recombinations: $k_{A \in \{S, K, I\}} \gg k_{B \in
\{\cleavage, \condensation\}}$.
The propensity function takes different forms depending on whether the
reaction is unimolecular or bimolecular, following the formulation of \citet{Gillespie:2007}. 
For unimolecular reactions, such as $I$ and $K$-reductions, and cleavages, $a_j$
takes the form $a_j(\sv) = c_j \sv_{x_1}$ where $\sv_{x_1}$ is the number of
copies of the reaction's substrate $x_1$ in $\sv$.
For bimolecular reactions like $S$-reductions and condensation with substrate
$x_1$ and reactant $x_2$, it takes the form $a_j(\sv) = c_j \sv_{x_1}
\sv_{x_2}$ if $x_1 \neq x_2$ and the form $a_j(\sv) = c_j \frac{1}{2} \sv_{x_1}
(\sv_{x_1} - 1)$ if $x_1 = x_2$, where $\sv_{x_1}$ and $\sv_{x_2}$ are the
number of expressions of $x_1$ and $x_2$, respectively.
In the case of unimolecular reactions, $c_j$ is equal to a reaction
rate constant $k_J$ where $J\in\{\cleavage,\reduction_K,\reduction_I\}$ is the
type of reaction $j$, whereas for bimolecular reactions $c_j = k_J/\Omega$ if
$x_1 \neq x_2$ and $2 k_J/\Omega$ if $x_1=x_2$, where $\Omega$ is the volume of
the simulated container~\citep{Gillespie:2007} and $J\in\{\condensation, \reduction_S\}$.

Then, the probability of the next reaction being $j$ is
$p(j)=a_j(\sv)/a_0$ where $a_0=\sum_j{a_j(\sv)}$, and the time
interval until its occurrence is distributed as an exponential distribution
with parameter $a_0$.
To sample from this process, we follow the direct method~\citep{Gillespie:1977}.
For efficiency reasons, we factorize the reaction probability by the kind
of the next reaction $J$ being a condensation ($\condensation$), a cleavage
($\cleavage$), or a reduction 
($\reduction = \reduction_S \cup \reduction_K \cup \reduction_I$), where 
$\{\reduction_Z\}_{Z \in \{S,K,I\}}$ stands for the set of all possible $Z$-reductions:

\begin{equation}
    p(j) = \sum_{J\in\{\condensation, \cleavage, \reduction\}}{p(j|J)p(J)},
\end{equation}
where $p(J) = a_J(\sv)/a_0(\sv)$, $a_0(\sv) =\sum_{J\in\{\condensation, \cleavage, \reduction\}}{a_J(\sv)}$, and $a_J = \sum_{j \in J}{a_j(\sv)}$.

For cleavage and condensation reactions, $a_J$ takes a simpler form that 
can be efficiently computed by keeping track of the total number of expressions $\sum_x\sv_x$, whereas 
for reduce reactions we must explicitly sum over all of reactions:
\begin{align}
    a_\cleavage(\sv) 
                      &= k_\cleavage \left(\left(\sum_x\sv_x\right) - \sv_S - \sv_K - \sv_I\right),
    \label{eq:propensity_start}\\
    a_\condensation(\sv) 
                         &= \frac{k_\condensation}{\Omega} \sum_x\sv_x \left( \sum_{x'}\sv_{x'} - 1 \right),\\
    a_\reduction(\sv) &=\sum_{j \in A_I}{k_{I} \sv_{j_1}} + \sum_{j \in A_K}{k_{K}\sv_{j_1}} + \sum_{j \in A_S}{\frac{k_{S}}{\Omega}\sv_{j_1}\sv_{j_2}},
                 \label{eq:propensity_end}
\end{align}
where $j_1$ is the substrate of the reduction reaction $j$ and $j_2$ is the reactant in the case of reducing an $S$ combinator. 

Thus, to sample a reaction, we first sample the reaction type $J$.
If it is a cleavage, then we sample one expression according to its concentration,
and cleave it into two subexpressions by dividing it at the root.
If it is a condensation, then we sample two expressions according to their concentration
(the second, after removing one element of the first one), and combine them
through the application operator.
Finally, if it is a reduction, then we sample one reduce reaction from the
space of all possible reduce reactions, with probability proportional to its
propensity.
In practice, we just need to compute all
possible reductions involving the expressions that are present in the system's
state\footnote{Some expressions can participate in a very large number
    of reductions, considerably slowing the simulation. For this reason, the 
system is currently limited to compute up to 10 reductions per expression in no special order.}.
The complete algorithm describing the temporal evolution of our system is
summarized on Algorithm \ref{al:reactor}\ifdefined\journal\footnote{We make 
available the code to simulate Combinatory Chemistry in \url{https://github.com/germank/combinatory-chemistry}.}\fi.

\ifinline{\algreactor}

\section{Emergent Structures}

Having described the dynamics of Combinatory Chemistry, we now turn to discuss
how we can characterize emergent structures in this system. 
For this, we first discuss how autocatalytic sets can be applied for this
purpose.
Second, we observe that this formalism may not completely account for some
emergent structures of interest, and thus, we propose to instead track reactant
consumption rates as a proxy metric to uncover the presence of these structures.
Finally, we enrich this metric to detect consumption levels that are above
chance levels.

\subsection{Autocatalytic Sets}

Self-organized order in complex systems is hypothesized to be driven by the
existence of attracting states in the system's dynamics~\citep{Walker:1966,
Wuensche:etal:1992, Kauffman:1993}.
Autocatalytic sets~\citep{Kauffman:1993} were first introduced by Stuart
Kauffman in 1971 as one type of such attractors that could help explaining the
emergence of life in chemical networks.
(See \citet{Hordijk:2019} for a comprehensive historical review on the topic.)
Related notions are the concept of autopoiesis \citep{Varela:Maturana:1973},
and the hypercycle model \citep{Eigen:1978}.

Autocatalytic sets (AS) are reaction networks that perpetuate in time by
relying on a network of catalyzed reactions, where each reactant and catalyst
of a reaction is either produced by at least some reaction in the network, or
it is freely available in the environment.
This notion was later formalized in mathematical form
\citep{Hordijk:Steel:2004,Hordijk:etal:2015} with the name of Reflexively
Autocatalytic Food-generated sets (RAFs).
Particularly, a Chemical Reaction System (CRS) is first defined to denote the
set of possible molecules, the set of possible reactions, and a catalysis set
indicating which reactions are catalyzed by which molecules.
Furthermore, a set of freely available molecules in the environment, called the
“food set”, is assumed to exist.
Then, anautocatalytic set (or RAF set) $\mathcal{S}$ of a CRS with associated food
set $F$ is a subset of reactions, which is:

\begin{enumerate}
\item \emph{reflexively autocatalytic} (RA): each reaction $r \in \mathcal{S}$
is catalyzed by at least one molecule that is either present in $F$ or can be
formed from $F$ by using a series of reactions in $\mathcal{S}$ itself. \label{it:ra}
\item \emph{food-generated} (F): each reactant of each reaction in
$\mathcal{S}$ is either present in $F$ or can be formed by using a series of
reactions from $\mathcal{S}$ itself. \label{it:f}
\end{enumerate}

\subsection{Metabolic Structures in Combinatory Chemistry}
\label{sec:autocatalysis_cc}

In Combinatory Chemistry, all reducing reactions take precedence over random
condensations and cleavages, and thus, they proceed at a higher rate than random
reactions without requiring to define a catalyst.
Therefore, we say that they are auto-catalyzed and note that they all
trivially satisfy condition \ref{it:ra}.
Thus, autocatalytic sets in this system are defined in terms of subsets of
reduce reactions in which every reactant is produced by a reduce reaction in
the set or is freely available in the environment (condition \ref{it:f}).
For example, if we assume that $A=(SII)$ is in the food set, Figure
\ref{fig:autocatalytic_set} shows a simple autocatalytic set associated with
the expression $(AA)=(SII(SII))$.
As shown, a chain of reduce reactions keeps the expression in a self-sustaining
loop:
When the formula is first reduced by the reaction $r_1$, a reactant $A$ is absorbed
from the environment and one $S$ combinator is released.
Over the following steps, two $I$ combinators are sequentially applied and
released back into the multiset $\mathcal{P}$, with the expression returning
back to its original form.
In other words, in one cycle the expression $(AA)$ absorbs one copy of $A$ from
the environment releasing back into it the elementary components obtained as
by-products of the reactions.
We refer to this process as a metabolic cycle because of its strong resemblance
to its natural counterpart.
For convenience, we write the just described cycle as $(AA) + A \reducesto (AA) + \phi(A)$,
where $\phi$ is a convenience function that allows us to succinctly represent the decomposition
of $A$ as elementary combinators by mapping its argument into $n_S S
+ n_K K + n_I I$ where $n_S$, $n_K$ and $n_I$ stand for the number of $S$, $K$
and $I$ combinators in $A$, and the $\reducesto$ symbol indicates that there
exists a pathway of reduction reactions from the reactives in the left hand
side to the products in the right hand side.

It should be noted that there are (infinitely) many possible pathways, some of
them not necessarily closing the loop. 
For instance, instead of reducing $(SII(I(SII))$ through the only available $I$
reduction ($r_4$ in Figure \ref{fig:autocatalytic_set}), it could also be
possible to reduce the full expression by applying the $S$ reduction rule as
long as at least one copy of $(I(SII))$ is present in $\mathcal{P}$, yielding
$(I(I(SII))(I(I(SII))))$ as a result.
We could continue reducing this expression, first through the two outermost $I$
combinators, thus obtaining $(SII(I(I(SII))))$, and then though the first $S$
combinator, provided that a copy of of $(I(I(SII))$ is present in
$\mathcal{P}$, obtaining $(I(I(I(SII)))(I(I(I(SII)))))$ as a result.
This outermost-first reduction order could be followed ad infinitum, stacking
ever more $I$ combinators within the expression.
Nonetheless, this also has a cost, as there must be copies of reactants
$(I(SII))$, $(I(I(SII)))$, $\dots$, $(I(\dots(I(SII))))$ in the multiset
$\mathcal{P}$ for these reactions to occur, and longer expressions are normally
scarcer.
For this reason, when we talk about the metabolic cycle we usually refer to the
``least effort'' cycle, in which $I$ and $K$ combinators are reduced before $S$
ones, and where $S$ combinators with shorter or naturally more frequent
expressions in the third argument (the reactant) are reduced before longer
ones.

\ifinline{
\begin{figure}[htpb]
\centering{
    \resizebox{240pt}{!}{
    \def\svgwidth{350pt}
    \input{figures/quine.pdf_tex}
    }
}
\caption{$r_1$--$r_5$ form an autocatalytic
set, granted that $(SII)$ belongs to the food set. $(SII(SII))$'s metabolic cycle
starts with $r_1$ reducing the $S$ combinator, while taking $(SII)$ as
reactant.
Then, the cycle is completed by the reduction of the two identity combinators,
in any of the possible orders.} 
\label{fig:autocatalytic_set}
\end{figure}
}

While autocatalytic sets provide a compelling formalism to study emergent
organization in Artificial Chemistry, they also leave some blind spots for
detecting emergent structures of interest.
Such is the case for \textbf{recursively growing} metabolisms.
Consider, for instance, $e = (S(SI)I(S(SI)I))$. 
This expression is composed of two copies of $A=(S(SI)I)$ applied to itself
$(AA)$. 
As shown in Figure \ref{fig:recursive}, during its metabolic cycle, it will
consume two copies of the element $A$, metabolizing one to perform its
computation, and appending the other one to itself, thus $(AA)  +
2A\reducesto (A(AA)) + \phi(A)$.
As time proceeds, the same computation will occur recursively, thus
$(A(AA)) + 2A \reducesto (A(A(AA))) + \phi(A)$, and so on.
While this particular behavior cannot be detected through autocatalytic sets,
because the resulting expression is not exactly equal to the original one,
it still involves a structure that preserves in time its functionality.

\ifinline{
\begin{figure}[htb]
    \centering
    \resizebox{170pt}{!}{
    \def\svgwidth{240pt}
    \input{figures/recursive.pdf_tex}
    }
\caption{One of the possible pathways in the reduction of the tail-recursive structure
$(AA)$ with $A=(S(SI)I)$. It appends one $A$ to itself by metabolizing another copy
absorbed from the environment.} 
\label{fig:recursive}
\end{figure}
}

Moreover, while the concept of autocatalytic set captures both patterns that
perpetuate themselves in time and patterns that also multiply their numbers, it
does not explicitly differentiate between them.
A pattern with a metabolic cycle of the form $AA + A \reducesto AA +
\phi(A)$ (as in Figure \ref{fig:autocatalytic_set}) keeps its own structure in
time by metabolizing one $A$ in the food set, but it does not self-reproduce. 
We call such patterns \textbf{simple autopoietic}~\citep{Varela:Maturana:1973}.
In contrast, for a pattern to be \textbf{self-reproducing} it must create
copies of itself that are later released as new expressions in the environment.
For instance, consider a metabolic cycle in Figure \ref{fig:self-reproducing}
with the form $(AA) + 3A \reducesto 2(AA) + \phi(A)$.
This structure creates a copy of itself from 2 freely available units of $A$
and metabolizes a third one to carry out the process.

\ifinline{
\begin{figure}[htpb]
\centering{
    \resizebox{200pt}{!}{
    \def\svgwidth{250pt}
    \input{figures/self-reproducing.pdf_tex}
    }
}
\caption{Metabolic cycle (showing one of the possible pathways) of a self-reproducing structure that emerges from the dynamics of Combinatory Chemistry.
    Starting from $(AA)$, where $A=(SI(S(SK)I))$, it acquires three copies of
    $A$ from its environment and uses two to create a copy of itself,
    metabolizing the third one to carry out the process.
}
\label{fig:self-reproducing}
\end{figure}
}

%
%
%

\subsection{Metrics for detecting emergence}
\label{sec:metric}

All structures identified in the previous section have in common the need
to absorb reactants from the environment to preserve themselves in
homeostasis.
Furthermore, because they follow a cyclical process, they will continuously 
consume the same types of reactants.
Thus, we propose tracking \emph{reactants consumption} as a proxy metric for
the emergence of structures.
In other words, we note that the only operation that allows an expression to
incorporate a reactant into its own body is the reduction of the $S$
combinator, and thus we count the number of reactants $x$ consumed by expressions
of the form $\alpha(Sfgx)\beta$ in the time interval [$t:t+\delta)$ normalized 
by the total number of reactants consumed in the same interval, denoting it
as $O_{t:t+\delta}(x)$, or simply $O(x)$.

Indeed, this metric was used on a previous version of this
work~\citep{Kruszewski:Mikolov:2020} to detect emergent structures.
Yet, it has the problem that it is also sensitive to reactants that are
consumed at high rates just because they are very frequent, and as a consequence,
expressions containing them as a third argument to an $S$ combinator are
expected to be common as well, inflating this metric for uninteresting reactions.

For this reason, we propose to use the (positive) \emph{pointwise
information}\footnote{The name and the formula are related to the pointwise
mutual information (PMI) metric \citep{Church:Hanks:1990} that has been
extensively used in computational linguistics. 
PMI computes the log ratio between the empirical co-occurrence probability of
two events $x$ and $y$ with respect to their expected probability would these
two events be independent, as given by the product of the marginals.
Like PMI, the proposed metric computes the log ratio between observed odds and
chance-driven ones.} $I(x)$ of having observed a consumption rate $O(x)$ for a
reactant $x$ with respect to its consumption rate should the process be driven
by chance only, $R(x)$:

\begin{equation}
    I(x)~\definition~\max\left(\log\frac{O(x)}{R(x)}, 0\right)
    \label{eq:information}
\end{equation}

We define $R(x)$ as the relative frequency of any expression on a hypothetical
process where no reduce reactions are present, but instead only random mixing
given by random collisions and cleavages.
To model this process, we follow the formulation of the reaction kinetic 
equations given by \citet{Fellermann:etal:2017}:

\begin{align}
    \od{\sv_x}{t} & =  \frac{c_\cleavage}{c_0} \left(\sum_{\substack{y,z\\(xy)=z\\(yx)=z}}{\sv_z} - \sum_{\substack{y,z\\(yz)=x}}{\sv_x}\right)\nonumber\\
                  &  + \frac{c_\condensation}{c_0} \left(\sum_{\substack{y,z\\(yz)=x}}{\sv_y \sv_z} - \sum_{\substack{y,z\\(xy)=z\\(yx)=z}}{\sv_x \sv_y} \right),
    \label{eq:random_kinetics}
\end{align}
where, $c_\cleavage=k_\cleavage$, $c_\condensation={k_\condensation}/{\Omega}$, and $c_0=a_0(\sv)$ is the partition function, and define $R(x)$ as the normalized equilibrium concentration
$\sv^*_x$ of expression $x$ under these random kinetics:
\begin{equation}
    R(x)~\definition~\frac{\sv^*_x}{\sum_x{\sv^*_x}}.
\end{equation}
At equilibrium, Equation \ref{eq:random_kinetics} admits the following solution:
\begin{equation}
    \sv^*_x = \clcoratio\me^{-\cb|x|},
    \label{eq:equilibrium}
\end{equation}
where $|x|$ is the length of expression $x$ and $\cb$ is a constant that 
depends on the boundary conditions.
In particular, we ask for the initial mass of the system, represented by the
dimensionless constant $\mass$, to be conserved in the equilibrium
distribution by equating it to the total number of combinators:
\begin{equation}
    \sum_{x}{\sv^*_x |x|} = \mass,
    \label{eq:boundary}
\end{equation}
We can rewrite this equation
by first substituting $\sv_x^*$ according to Equation \ref{eq:equilibrium}, 
and then replacing the sum over expressions by a sum over lengths.
Noting that there are $\ck^n \cat{n-1}$ expressions of length $n$, where $\ck=3$
is the number of different possible combinators and $C_n$ is the $n$th Catalan 
number, this equation becomes: 
\begin{equation} 
    \clcoratio\sum_{n=1}^\infty{\ck^nC_{n-1}\me^{-\cb n}}n = \mass.
    \label{eq:mass_conservation}
\end{equation}
After solving this equation for $\cb$, we obtain $\cb = \log{\left(2 \ck + 2 \ck\sqrt{1 + \left(\clcoratio\right)^2\frac{1}{4\mass^2}}\right)}$.
    Next, we compute the normalization constant for the distribution $\sv^*$:
\begin{align}
\sum_x{\sv^*_x} = \clcoratio\sum_{n=1}^\infty{\ck^nC_{n-1}\me^{-\cb n}} = \frac{\cc_\cleavage}{2\cc_\condensation}\left(1 - \sqrt{1 - 4\ck\me^{-b}}\right),
    \label{eq:normalization}
\end{align}
completing the definiton of $I(x)$. We refer to the Appendix \ref{ap:kinetics}
for all the computations verifying these claims. 
In the following section, we show its effects experimentally. 

\section{Experiments and Discussion}

\subsection{Metrics}
We began by testing the effect of the proposed information metric on a system
with uniformly distributed $\mass=10^4$ $S$, $K$ and $I$ combinators simulated until
it reached time $T=1000$.
Candidate reactions are sampled by $10$ threads working simultaneously.
For all of our experiments, we used $k_\condensation=k_\cleavage=1$,
$k_K=k_I=10^4$, $k_S=10^6$ and fix the dimensionless constant
$\Omega=\mass$. 
We leave a thorough exploration of parameter values for future work.
With these parameters, the expressions associated with random kinetic
dynamics become $\cb = \log\left((2 +
\sqrt{5})\ck\right)$ and $R(x) = \frac{1}{2} (3 +
\sqrt{5})\left((2+\sqrt{5})k\right)^{-|x|}$.
All curves are smoothed through locally averaging
every data point at time $t$ with those in the interval $[t-10, t+10]$.

We first present the raw consumption rates $O(x)$, which are displayed in Figure
\ref{fig:consumption-rates-base}. 
As shown, some of the most frequently consumed reactants include the atomic
combinators $S$, $K$ and $I$, and some of their binary compositions.
Binary reactants such as $(KI)$ and atomic ones such as $I$ do not form
part of any stable structure, and the expressions consuming them are produced
by chance.
Yet, they are used with considerable frequency because $S$ combinators are
more likely to be applied to shorter arguments than longer ones.
For this reason, the consumption of $I$ is considerable higher than the
consumption of $(KI)$.
Yet, even though by the same argument the consumption rate of $A=(SII)$ should
be below binary reactants, self-organisation into autopoietic patterns drives
the usage of this reactant above what would be expected if chance would be the
only force at play.
Indeed, the curve corresponding to the consumption of the reactant
$A=(SII)$ is associated with the autopoietic pattern $(SII(SII))$, composed of
two copies of this reactant, and a metabolic cycle of the form $(AA) + A
\reducesto (AA) + \phi(A)$, as shown on Figure 
\ref{fig:autocatalytic_set}.
Nonetheless, $(S(SI)I)$, used by the structure in Figure
\ref{fig:recursive}, is cramped at the bottom with the binary reactants
in the consumption rate plot.

When applying the proposed information metric $I(x)$ the curves corresponding
to emerging structures become featured at the top of the graph whereas all
other reactants that are mostly driven by random generation are pushed to the
bottom, as shown on Figure \ref{fig:consumption-rates-info}.
Therefore, this experiment showcases the usefulness of this metric in separating
consumption rates propelled by emergent structures from uninteresting
fortuitous ones.

\ifinline{\figweighting}

\subsection{Emergent metabolic structures}

\ifinline{\figglobalmetrics}

For the next part, we studied some of the emergent structures in a system
initialized with a \emph{tabula rasa} state consisting of $\mass=10^6$ evenly
distributed $S$, $K$ and $I$ combinators simulated for $1000$ units of time.

In an earlier version of this work~\citep{Kruszewski:Mikolov:2020}, we had used
a supplementary mechanism called reactant assemblage, through which we ``fed''
emerging structures with their required reactants to allow them to be spotted
through sheer reactant consumption rates on a relatively small system with only
$10$k combinators. 
Here, we simplified the model and dropped the need for this mechanism, thanks
both to simulating a larger system with $1$M combinators and to the usage of
the re-weighting metric presented in Section \ref{sec:metric}, which allowed us
to spot emergent complex structures even even at very low levels of reactant
consumption rates.

\ifinline{\figemergence}

We began by analyzing some general metrics on one given run of the system.
First, and in agreement with previous work~\citep{Meyer:etal:2009}, we find
that there is a tendency for the system to create increasingly longer
expressions, as shown by the length of the largest expression in the system
(Figure \ref{fig:max-length}).
We also count the number of distinct expressions present
at in the system at any given time, which we display on Figure
\ref{fig:diversity}.
As can be seen, diversity explodes at the beginning driven by
the random recombination of the elementary combinators, peaking
very early on, then decreasing fast at first, and then slower after about
time $200$.
This behavior is consistent with a system that self-organizes into
attracting states dominated by fewer, but more frequent expressions.
Again, this result agrees with previous work that has shown that diversity
decreases over time on a number of ACs~\citep{Dittrich:Banzhaf:1998,
Banzhaf:etal:1996,Dittrich:2001}. 
However, we note that in contrast with many of these systems that were
initialized with random elements, thus, maximizing diversity from the start, here
the initial diversity is an emergent property of the system dynamics as it is
only initialized with $3$ different elements, namely, the singleton expressions
$S$, $K$ and $I$.
Next, we note that the proportion of reducing vs random recombination
reactions is an emergent property of the system which depends on the number of
reducible expressions that are present in its state at any given time.
As can be seen in Figure \ref{fig:p-reduce}, this rate, which necessarily
starts at $0$, increases sharply in the beginning, reaching slightly more than
30\%.
Then, it starts to slowly decrease, but remains always above 25\% during
the studied period.

%
However, it is unclear from these results whether there are  emergent complex
structures that act as attractors, or if a different explanation for these
outcomes is at play.
To answer this question, we turned to the reactant consumption rates weighted
by Equation \ref{eq:information} to detect whether specific reactants were
more prominently used by some emergent structures.

Results are shown in Figure \ref{fig:consumption} 
for a few selected reactants that highlight the emergence of different types of
structures, including simple autopoietic, recursively growing, and
self-reproducing ones.
Interestingly, they can emerge at different points in time, co-exist, or be
driven to extinction.
In parallel, Figure \ref{fig:reactants-count} show the number of copies of
these reactants available at each discrete time interval of unit length.

While there are (infinitely) many possible expressions that can consume a 
given type of reactant, only a few of them will correspond to emergent 
metabolisms. 
In general, we observed that expressions that consume any given reactant $A$ are
typically composed of multiple juxtaposed copies of this reactant in an
expression of the form $(AA)$. 
This is linked to the fact that in order to express recursive functions in
combinatory logic, the function (in this case denoted by $A$) must take itself
as its own argument, but this particularity also confirms the old adage: ``Tell
me what you eat and I will tell you what you are''. 

The first curve in Figure \ref{fig:consumption} (in the order of the
legend) correspond to the reactant $A=(SII)$, associated with the \emph{simple
autopoietic} structure of Figure \ref{fig:autocatalytic_set}.
The consumption rates are much more stabilized in comparison with the results reported
in Figure \ref{fig:consumption-rates-info}, which belonged to a system that was
$100$ times smaller.
Furthermore, the number of copies of this reactant decreases sharply at the beginning,
but then, intriguingly, they slowly increase again until reaching a
concentration of about $200$ units in total.

The second reactant in the plot, $(AA)$ (where $A=(SII)$) is not actually
consumed by a metabolism. 
Instead, it is consumed by one of the two possible reductions of the expression 
$(A(AA)) = (SII(SII(SII))$, which reduces to $(AA(AA)) = (SII(SII)(SII(SII)))$.
This new structure is yet another autopoietic structure that is composed
of two copies of $(AA)=(SII(SII))$ which persist in time by consuming the
$(SII)$ reactant independently of each other.

The next three reactants in Figure \ref{fig:consumption} correspond to
\emph{recursively growing} structures. 
The first one uses the reactant $A=(S(SI)I)$, and follows a right-branching
cycle that linearly increases the size of the structure: $(AA) + 2A
\reducesto (A(AA)) + \phi(A)$ (Figure \ref{fig:recursive}).
The second one, with reactant $A=(S(SII)I)$, also grows recursively, although
with a left-branching structure: $(AA) + 2A \reducesto (AAA) + \phi(A)$.
Third, there is the reactant $A=(S(SSI)K)$, which is associated with a 
binary-branching recursive structure\footnote{See Appendix \ref{ap:derivations}
for more details on these derivations.}.
%

Finally, the curve of the reactant $A=(SI(S(SK)I))$ corresponds to the
emergence of a \emph{self-reproducing} structure, following a
cycle of the form $(AA)  + 3A \reducesto 2(AA) + \phi(A)$, thus
duplicating itself after metabolizing one copy of the reactant in the process.
Interestingly, this structure emerges not only through the effect of random 
recombination, but also thanks to self-organization.
Instead of being a product of a random combination of two copies of the
reactant $A$, it often emerges after the condensation of other reactants, such
as $(S(SI(SI))(SI))$ and $(S(SK)I)$, $(SI(SI(SI)))$ and $(S(SK)I)$, or $(SII)$
and $(SI(S(SK)I))$, among other possibilities that induce a chain of reduction
reactions that result in producing at least one copy of $(AA)$.

It is worth noting that one cycle of this structure's metabolism requires
three copies of the reactant. 
When there are none in the environment, the structure cannot proceed with its
metabolism and this structure is vulnerable to being cleaved or of being
condensed with an expression that will cause it to stop functioning normally.
Because of the rare supply of its six-combinator-long reactant, plus the fact
that all existing structures compete with each other, they fall into extinction
at about time $t=200$, although they make a short come-back at time $t=400$,
when only two reproduction cycles were completed before falling back into
oblivion.
Nonetheless, we speculate that in larger systems the population will recover
from periods of resource scarcity by following the periodic dynamics 
originally proposed by~\citet{Lotka:1910}, and thus allowing these structures to
perpetuate in time.

Notably, recursive structures also experience low resource conditions
starting at time $t=200$. However, they might be able to cope with conditions
of low resources more effectively because of their repetitive structure allow
them to be cleaved and still conserve their function.
For instance, $A(AA) \reducer A + (AA)$ still leaves a functioning $(AA)$
structure.
When new copies of $A$ become available either through the random condensation of combinators released by every computed reduction
or by some other process, they can consume them and grow back again.

\subsection{Other Bases}
\label{sec:other-bases}

Thus far, we have explored emergent structures on systems composed of $S$, $K$, 
and $I$ combinators.
Because one of the main goals of this work is linking the emergence of
metabolic structures to the core computational properties of the underlying
chemistry, we explored other possible (smaller) bases.

First of all, we note that using $K$ or $I$ combinators by themselves would not
produce any meaningful structure, neither would a combination thereof.
Any expression formed out of these combinators can only decay into binary 
expressions at most.
In the case of the $S$ combinator, while it still allows expressions that can
be reduced infinitely, such as for instance, $(SSS(SSS)(SSS))$, there are no
expressions which do so by continually consuming the same reactant.
This is because, as shown by \citet{Waldmann:2000}, there are no expressions
$X$ composed only of the $S$ combinator such that $X \reducesto
(\alpha{}X\beta)$.
Therefore, auto-catalytic sets are not possible in this environment.
For all of these reasons, it is not meaningful to look at simulations
containing either one of the $S$, $K$ or $I$ combinators alone.

The only two possible remaining subsets of combinators are the pairs $S-I$ and
$S-K$, with only the latter being Turing-complete.
We ran again experiments in which only one pair of combinators was present.
We used $10^4$ combinators for the $S-I$ basis, and $10^6$ for the $S-K$ one.
Figure \ref{fig:base-SI} show the information traces for the reactant $(SII)$,
consumed by the simple autopoietic structure in Figure
\ref{fig:autocatalytic_set}, and $(S(SI)I)$, consumed by the structure in
Figure \ref{fig:recursive}, corroborating that these structures also emerge 
when only $S$ and $I$ combinators are present.
In Figure \ref{fig:base-SK}, we also note that a homologous autopoietic
structure emerges with $S-K$ basis, as shown by the consumption of the
reactant $A=(S(SK)(SKK))$.
This structure has a metabolic cycle of the form $(AA) + 3A \reducesto
(AA) + 2A + \phi(A)$, thus needing to absorb 2 extra copies of the reactant to
perform the computation, even if they are later released unchanged.
Additionally, recursively growing structures can be spotted in the $S-K$ base, as shown
by the consumption of $A=(S(SSK))$, which is associated with a metabolic
cycle of the form $(AAA) + 3A + AA \reducesto (SSKA(AAA)) + 4S + K + AA$,
thus growing and incorporating $SSK$ as a prefix in the process.
Interestingly, $AA$ is absorbed and released intact, which could be construed
as an emergent catalyst for the reaction: Even though we can interpret
each reduce reaction to be auto-catalyzed, reaction chains can have emergent
properties, such as in this case, where a reactant is just used to complete the
metabolic cycle and then released.

Finally, we note that while we have not yet witnessed an emergent
self-reproducing structure using the $S-K$ only, it can in fact be represented
with the expression $(AA)$ where $A=(S(SK)(S(SK)(SKK)))$, and having as
metabolic cycle $(AA) + 5A \reducesto 2 (AA) + A + KA + 5S + 3K$.
However, finding it requires discovering a considerably longer expression than
the one in the $SKI$ basis and consumes much longer reactants, which might
explain why we have not yet found them to emerge in our simulations at the current
scale.
Nonetheless, it is worth noting that in the incomplete $S-I$ basis it would
not be possible to represent a self-reproducing metabolic cycle of the form $X
\reducesto 2X + \dots$ for any $X$ because this would ask for a
reaction capable of producing as a by-product an arbitrarily long expression
that reduces to $X$ at some point during the cycle.
Yet, $S$ and $I$ combinators only have themselves as by-products of their
respective reactions and cannot fulfill this requisite.

\ifinline{\figotherbases}

\section{Conclusions}

We have introduced Combinatory Chemistry, an Algorithmic Artificial Chemistry
based on Combinatory Logic. 
Even though it has relatively simple dynamics, it gives rise to a wide range of
autopoietic structures, including recursively growing and self-reproducing
ones.
These structures  feature reaction cycles that bear a striking resemblance to
natural metabolisms.
All of them take the form of recursive algorithms that continually consume
specific resources, incorporating them into their structure and decomposing
them to perform their function.
Thanks to Combinatory Logic being Turing-complete, the presented system can
theoretically represent patterns of arbitrary complexity.
Furthermore, we have argued that in the context of the $SKI$ basis, this
computational universality property is both necessary and sufficient to
represent self-reproducing patterns, while also have shown them to emerge at
least in the case when all three combinators are present.
On the other hand, a non-universal basis consisting only of $S$ and $I$ combinators
can still give rise to simple autopoietic and recursively-growing structures.

The proposed system does not need to start from a random set of initial
expressions to kick-start diversity.
Instead, this initial diversity is the product of the system's own dynamics, as
it is only initialized with elementary combinators.
In this way, we can expect that this first burst of diversity is not just a 
one-off event, but it is deeply embedded into the mechanics of the system, 
possibly allowing it to keep on developing novel structures continually.

To conclude, we have introduced a simple model of emergent complexity in which
self-reproduction emerges autonomously from the system's own dynamics.
In the future, we will seek to apply it to explain the emergence of
evolvability, one of the central questions in Artificial Life.
While many challenges lie ahead, we believe that the simplicity of the model,
the encouraging results presently obtained, and the creativity obtained from
balancing computation with random recombination to search for new forms,
leaves it in good standing to tackle this challenge.


\footnotesize
\printbibliography

\appendix
\section{Random kinetics derivations}
\label{ap:kinetics}

\subsection{Equilibrium distribution}
Here we show that Equation \ref{eq:equilibrium} corresponds to the equilibrium distribution
of the process defined by Equation \ref{eq:random_kinetics}. 
Plugging Equation \ref{eq:equilibrium} into Equation \ref{eq:random_kinetics}
we obtain:

\begin{align}
    \od{\sv^*_x}{t} & =  \frac{\cc_\cleavage}{c_0} 
    \left(\sum_{\substack{(xy)=z\\(yx)=z}}{\frac{\cc_\cleavage}{\cc_\condensation}\me^{-\cb|z|}} - \sum_{(yz)=x}{\frac{\cc_\cleavage}{\cc_\condensation}\me^{-\cb|x|}}\right)\nonumber\\
                &  + \frac{\cc_\condensation}{c_0} 
                \left(\sum_{(yz)=x}{\frac{\cc^2_\cleavage}{\cc^2_\condensation}\me^{-\cb(|y|+|z|)}} - \sum_{\substack{(xy)=z\\(yx)=z}}{\frac{\cc^2_\cleavage}{\cc^2_\condensation}\me^{-\cb(|x|+|y|)}} \right).
\end{align}
Noting that if $z = (xy)$, then $|z| = |x| + |y|$, that an expression
\hbox{$x=(x_l x_r)$} can only be cleaved into $x_l$ and $x_r$, while vice versa,
only the condensation of $x_l$ and $x_r$ can form $x$, and that there are
$\ck^nC_{n-1}$ possible expressions of length $n$ (where $C_n$ stands for the
$n$th catalan number and $\ck=3$ is the number of combinators), and that
we must sum twice the factors corresponding to expressions $z$ that can be 
formed either as $(xy)$ or as $(yx)$, then we have:

\begin{align}
    \od{\sv_x}{t} & =  \frac{1}{c_0} 
    \left(\left(\sum_{n=1}^\infty{\frac{\cc^2_\cleavage}{\cc_\condensation}2\times \ck^n C_{n-1} \me^{-\cb(|x|+n)}}\right) - {\frac{\cc^2_\cleavage}{\cc_\condensation}\me^{-\cb|x|}}\right)\nonumber\\
                &  + \frac{1}{c_0} 
                \left({\frac{\cc^2_\cleavage}{\cc_\condensation}\me^{-\cb(|x|)}} - \left(\sum_{n=1}^\infty{\frac{\cc^2_\cleavage}{\cc_\condensation}2\times \ck^n C_{n-1}\me^{-\cb(|x|+n)}}\right)\right).
\end{align}
This expression evaluates to $0$ as long as the series converges, which we can assess
using the ratio test, and the identity $C_{n+1} = \frac{2(2n+1)}{n+2}C_n$:

\begin{equation}
    \lim_{n \to \infty}{\frac{\ck^{n+1}C_n\exp\left(-\cb(|x|+n+1)\right)}{\ck^{n}C_{n-1}\exp\left(-\cb(|x|+n)\right)}} =  \frac{2\ck(2n - 1)}{n+1}e^{-\cb} < 1.
\end{equation}
Thus, if $\cb > \log (4\ck)$, then Equation \ref{eq:equilibrium} defines the equilibrium
distribution.

\subsection{Boundary conditions}
Here, we derive the value of $\cb$ by solving Equation \ref{eq:mass_conservation}.
We start by showing that the following identity holds:
\begin{equation}
    \sum_{n=1}^\infty{\ck^nC_{n-1}\me^{-\cb n}}n = \frac{\ck \me^{-\cb}}{\sqrt{1 - 4\ck{}\me^{-\cb}}}.
\end{equation}
Rearranging the terms of the series, we have:
\begin{align}
    \sum_{n=1}^\infty{\ck^nC_{n-1}\me^{-\cb n}}n &= \sum_{n=1}^\infty{C_{n-1}\me^{-(\cb - \log k) n}}n  = \sum_{n=1}^\infty{C_{n-1}a^nn}\nonumber\\
                                                 &=a\sum_{n=1}^\infty{C_{n-1}a^{n-1}n}
                                                 =a\sum_{m=0}^\infty{C_ma^m (m+1)},
                                                 \label{eq:full_series}
\end{align}
where $a=\me^{-(\cb-\log \ck)}$, $m = n-1$.
Then, using the definition of $C_n = \frac{1}{n+1}{2n \choose n}$, first, and the 
generating function for central binomial coefficients~\citep{Lehmer:1985}, second,
we obtain:
\begin{align}
    \sum_{m=0}^\infty{C_ma^m(m+1)} = \sum_{n=0}^\infty{{2m \choose m}a^m}=\frac{1}{\sqrt{1 - 4a}},
    \label{eq:binom_series}
\end{align}
with $|a| < 1/4$.
Finally, replacing Equation \ref{eq:binom_series} into \ref{eq:full_series} and expanding $a$, we have:
\begin{align}
    \sum_{n=1}^\infty{\ck^nC_{n-1}\me^{-\cb n}}n &= \me^{-(\cb-\log \ck)} \frac{1}{\sqrt{1 - 4\me^{-(\cb-\log \ck)}}} = \frac{k\me^{-\cb}}{\sqrt{1 - 4\ck\me^{-\cb}}},\label{eq:series1_sol}
\end{align}
with $b > \log{4\ck}$.
Next, we replace Equation \ref{eq:series1_sol} into \ref{eq:mass_conservation},
\begin{align}
\clcoratio\frac{k\me^{-\cb}}{\sqrt{1 - 4\ck\me^{-\cb}}} = \mass
\end{align}
and solve for $\cb$ to obtain:
\begin{align}
    \cb = \log{\left(2 \ck + 2 \ck\sqrt{1 + \left(\clcoratio\right)^2\frac{1}{4\mass^2}}\right)}.
\end{align}

\subsection{Normalizing constant}

Next, we compute the normalizing constant $\sum_x{\sv^*_x}$. We start by showing the derivation for the following identity: 
\begin{align}
    \sum_{n=1}^\infty{\ck^nC_{n-1}\me^{-\cb n}} &= \frac{1}{2}\left(1 - \sqrt{1 - 4\ck\me^{-\cb}}\right).
\end{align}
Using this time the generating function for Catalan numbers~\citep{Davis:2006},
\begin{equation}
    \sum_{n=0}^\infty{C_na^n} = \frac{1 - \sqrt{1 - 4a}}{2a},
\end{equation}
we follow an analogous argument to the one above:
\begin{align}
\sum_{n=1}^\infty{\ck^nC_{n-1}\me^{-\cb n}} &= \sum_{n=1}^\infty{C_{n-1}\me^{-(\cb - \log \ck)n}} = 
\sum_{n=1}^\infty{C_{n-1}a^n} = a \sum_{n=1}^\infty{C_{n-1}a^{n-1}} \nonumber\\
                                            &= a\sum_{m=0}^\infty{C_ma^m} = \frac{1 - \sqrt{1-4a}}{2} = \frac{1 - \sqrt{1 - 4\ck\me^{-\cb}}}{2}.
\end{align}
where $a=\me^{-(\cb-\log \ck)}$, $m = n-1$. Thus, using again that there are $\ck^nC_{n-1}$ expressions of length $n$:
\begin{align}
\sum_x{\sv^*_x} = \clcoratio\sum_{n=1}^\infty{\ck^nC_{n-1}\me^{-\cb n}} = \frac{\cc_\cleavage}{2\cc_\condensation}\left(1 - \sqrt{1 - 4\ck\me^{-\cb}}\right).
\end{align}

\section{Metabolic cycles}
\label{ap:derivations}

The following derivations show one of the possible pathways that each of the
described structures can undertake as they develop.
Whenever more than one reduction is possible, the ``least effort'' path is
followed, namely, $I$ and $K$ combinators are reduced first, and then $S$ 
combinators with the shortest reactant (i.e. third argument).
Also, note that every expression written as $((fx)(gy))$ can also be written
simply as $(fx(gy))$, a fact that we often make use of when applying an
$S$-reduction.

\subsection{Metabolic cycle of a simple autopoietic pattern}
Let $A=(SII)$. Then,

\begin{align*}
(\underline{AA}) + A &\reducer& ((IA)(IA)) + S\\
(\underline{IA}(IA)) & \reducer & (A(IA)) + I \\
(A(\underline{IA})) & \reducer & (AA) + I \\
\end{align*}

\subsection{Metabolic cycle of a right-branching recursively growing structure}

Let $A=(S(SI)I)$. Then,

\begin{align*}
(\underline{AA}) + A &\reducer& (SIA(IA)) + S\\
(SIA(\underline{IA})) & \reducer & SIAA + I\\
(\underline{SIAA}) + A & \reducer & (IA(AA)) + S\\
(IA(AA)) & \reducer & (A(AA)) + I \\
\end{align*}

\subsection{Metabolic cycle of a binary-branching structure}

Let $A=(S(SSI)K)$. Then $(AA)$ can follow the metabolic pathway:

\begin{align*}
(\underline{AA}) + A &\reducer& (SSIA(KA)) + S\\
\underline{SSIA}(KA) + A&\reducer& (SA(IA)(KA)) + S\\
(SA(\underline{IA})(KA)) &\reducer& (SAA(KA)) + I\\
(\underline{SAA(KA)}) + (KA) &\reducer& (A(KA)(A(KA))) + S
\end{align*}

Then each copy of $(A(KA))$ can be reduced as follows

\begin{align*}
(\underline{A(KA)}) + (KA) &\reducer& SSI(KA)(K(KA)) + S\\
(\underline{SSI(KA)}(K(KA)) + (KA) &\reducer& (S(KA)(I(KA))(K(KA))) + S\\
(S(KA)(\underline{I(KA)})(K(KA))) &\reducer& (S(KA)(KA)(K(KA))) + I\\
(\underline{S(KA)(KA)(K(KA))}) + (K(KA)) &\reducer& (KA(K(KA))(KA(K(KA)))) + S\\
(\underline{KA(K(KA))}(KA(K(KA)))) &\reducer& (A(KA(K(KA)))) + (K(KA)) + K\\
(A(\underline{KA(K(KA))})) &\reducer& (AA) + (K(KA)) + K
\end{align*}

Thus, the complete pathway can be summarized as $(AA) +2A + 5(KA) + (K(KA))
\reducesto AA(AA) + 4 (K(KA)) + 2\phi(A)$.

\subsection{Metabolic cycle of a self-reproducing expression}

Let $A=(SI(S(SK)I))$. Then,

\begin{align*}
(\underline{AA}) + A &\reducer& (IA(S(SK)IA)) + S \\
(\underline{IA}(S(SK)IA)) &\reducer& (A(S(SK)IA)) + I\\
(A(\underline{S(SK)IA}))  + A & \reducer & (A(SKA(IA))) + S\\
(A(SKA(\underline{IA}))) & \reducer & (A(SKAA)) + I \\
(A(\underline{SKAA})) & \reducer & (A(KA(AA))) + S \\
(A(\underline{KA(AA)})) & \reducer & (AA) + (AA) + K \\
\end{align*}

\subsection{Arrival of the self-reproducing expression}

In our simulations, we found that $(AA)$ with $A=(SI(S(SK)I))$ often emerged
from the condensation of two expressions leading to a chain of reactions that
resulted in $(AA)$. Here we show one simple path involving the condensation
of $(SI(SI))$ and $(S(SK)I)$ to produce $(SI(SI)(S(SK)I))$.
Let's call $B=(S(SK)I)$, and note that $A=(SIB)$. 
The reduction chain that leads to $(AA)$ proceeds as follows:

\begin{align*}
    (\underline{SI(SI)B}) + B  & \reducer & (IBA) + S\\
    (\underline{IB}A) & \reducer & (BA) + I\\
    (\underline{S(SK)IA}) + A & \reducer & (SKA(IA)) + S\\
    (SKA(\underline{IA})) & \reducer& (SKAA) + I\\
    (\underline{SKAA}) + A & \reducer & (KA(AA)) + S\\
    (\underline{KA(AA)}) & \reducer & AA + A + K
\end{align*}


\subsection{Metabolic cycle of a simple autopoietic pattern on the $S-K$ basis}
Let $A=(S(SK)(SKK))$. Then,

\begin{align*}
(\underline{AA}) + A &\reducer& ((SKA)(SKKA)) + S\\
(SKA(\underline{SKKA})) + A & \reducer & (SKA(KA(KA))) + S\\
(SKA(\underline{KA(KA)})) & \reducer & (SKAA) + KA + K \\
(\underline{SKAA}) + A & \reducer & (KA(AA)) + A\\
(\underline{KA(AA)}) & \reducer & A + (AA) + K\\
\end{align*}

\subsection{Metabolic cycle of a recursively-growing expression on the $S-K$ basis}

%

Let $A=(S(SSK))$. Then, 

\begin{align*}
    (\underline{AAA}) + A & \reducer & (SSKA(AA)) + S\\
    (\underline{SSKA}(AA)) + A & \reducer & (SA(KA)(AA)) + S\\
    (\underline{SA(KA)(AA)} + AA & \reducer & (A(AA)(KA(AA))) + S\\
    (A(AA)(\underline{KA(AA)})) & \reducer & (A(AA)A) + (AA) + K \\
    (\underline{A(AA)A}) + A & \reducer & (SSKA(AAA)) + S 
\end{align*}

\subsection{Metabolic cycle of a self-reproducing expression on the $S-K$ basis}

Let $A=(S(SK)(S(SK)(SKK)))$. Then,

\begin{align*}
    (\underline{AA}) + A &\reducer& (SKA(S(SK)(SKK)A)) + S \\
    (SKA(\underline{S(SK)(SKK)A})) + A & \reducer & (SKA(SKA(SKKA))) + S \\
    (SKA(SKA(\underline{SKKA}))) + A & \reducer & (SKA(SKA(KA(KA)))) + S \\
    (SKA(SKA(\underline{KA(KA)}))) & \reducer & (SKA(SKAA)) + K + KA \\
    (SKA(\underline{SKAA})) + A & \reducer & (SKA(KA(AA))) + S \\
    (SKA(\underline{KA(AA)})) & \reducer & (SKAA) + AA + K \\
    (\underline{SKAA}) + A & \reducer & (KA(AA)) + S \\
    (\underline{KA(AA)}) & \reducer & A + AA + K\\
\end{align*}

\pagebreak
\ifdefined\journal
\algreactor
\pagebreak

\begin{figure}[htpb]
\centering{
    \resizebox{240pt}{!}{
    \def\svgwidth{350pt}
    \input{figures/quine.pdf_tex}
    }
}
\caption{$r_1$--$r_5$ form an autocatalytic
set, granted that $(SII)$ belongs to the food set. $(SII(SII))$'s metabolic cycle
starts with $r_1$ reducing the $S$ combinator, while taking $(SII)$ as
reactant.
Then, the cycle is completed by the reduction of the two identity combinators,
in any of the possible orders.} 
\label{fig:autocatalytic_set}
\end{figure}

\pagebreak

\begin{figure}[htb]
    \centering
    \resizebox{170pt}{!}{
    \def\svgwidth{240pt}
    \input{figures/recursive.pdf_tex}
    }
\caption{One of the possible pathways in the reduction of the tail-recursive structure
$(AA)$ with $A=(S(SI)I)$. It appends one $A$ to itself by metabolizing another copy
absorbed from the environment.} 
\label{fig:recursive}
\end{figure}

\pagebreak

\begin{figure}[htpb]
\centering{
    \resizebox{200pt}{!}{
    \def\svgwidth{250pt}
    \input{figures/self-reproducing.pdf_tex}
    }
}
\caption{Metabolic cycle (showing one of the possible pathways) of a self-reproducing structure that emerges from the dynamics of Combinatory Chemistry.
    Starting from $(AA)$, where $A=(SI(S(SK)I))$, it acquires three copies of
    $A$ from its environment and uses two to create a copy of itself,
    metabolizing the third one to carry out the process.
}
\label{fig:self-reproducing}
\end{figure}

\pagebreak
\figweighting
\pagebreak
\figglobalmetrics
\pagebreak
\figemergence
\pagebreak
\figotherbases
\fi

\end{document}